\def\streamavcombined{1}
\documentclass{article}
\usepackage{iclr2027_conference,times}
\usepackage{graphicx}
\usepackage{amsmath}
\usepackage{booktabs}
\usepackage{pifont}
\usepackage{array}
\usepackage{xspace}
\usepackage{adjustbox}
\usepackage{multirow}
\usepackage{xcolor}
\usepackage{float}
\usepackage{listings}
\usepackage[most]{tcolorbox}
\usepackage{hyperref}
\usepackage{xurl}
\usepackage[capitalize]{cleveref}

\newcommand{\cmark}{\ding{51}}
\newcommand{\xmark}{\ding{55}}
\newcommand{\shuchen}[1]{\textcolor[rgb]{0.961,0.137,0}{{#1}}}

\makeatletter
\DeclareRobustCommand\onedot{\futurelet\@let@token\@onedot}
\def\@onedot{\ifx\@let@token.\else.\null\fi\xspace}

\def\eg{\emph{e.g}\onedot} \def\Eg{\emph{E.g}\onedot}
\def\ie{\emph{i.e}\onedot} \def\Ie{\emph{I.e}\onedot}
\def\cf{\emph{c.f}\onedot} \def\Cf{\emph{C.f}\onedot}
\def\etc{\emph{etc}\onedot}
\def\vs{\emph{vs}\onedot}
\def\wrt{w.r.t\onedot} \def\dof{d.o.f\onedot}
\def\etal{\emph{et al}\onedot}
\let\savedmaketitle\maketitle
\let\savediclr@maketitle\@maketitle
\let\savedthanks\thanks
\makeatother

\newtcblisting{promptbox}[1]{
  enhanced,
  breakable,
  listing only,
  colback=gray!4,
  colframe=black!55,
  boxrule=0.5pt,
  arc=1.5pt,
  left=4pt,
  right=4pt,
  top=3pt,
  bottom=3pt,
  title={#1},
  fonttitle=\bfseries\small,
  listing options={
    basicstyle=\ttfamily\scriptsize,
    breaklines=true,
    breakatwhitespace=true,
    columns=fullflexible,
    keepspaces=true,
    showstringspaces=false,
    literate={—}{{---}}1 {–}{{--}}1 {×}{{$\times$}}1
  }
}

\renewcommand{\topfraction}{0.9}
\renewcommand{\dbltopfraction}{0.9}
\renewcommand{\textfraction}{0.08}
\renewcommand{\floatpagefraction}{0.8}
\renewcommand{\dblfloatpagefraction}{0.85}
\iclrfinalcopy

\begin{document}

\ifdefined\streamavcombined
\else
\pdfminorversion=7
\documentclass{article}
\usepackage{iclr2027_conference,times}
\usepackage{graphicx}
\usepackage{amsmath}
\usepackage{booktabs}
\usepackage{pifont}
\usepackage{array}
\usepackage{xspace}
\usepackage{adjustbox}
\usepackage{multirow}
\usepackage{flafter}
\usepackage{xcolor}
\usepackage{hyperref}
\usepackage{xurl}
\usepackage[capitalize]{cleveref}

\newcommand{\cmark}{\ding{51}}
\newcommand{\xmark}{\ding{55}}
\newcommand{\shuchen}[1]{\textcolor[rgb]{0.961,0.137,0}{{#1}}}

\makeatletter
\DeclareRobustCommand\onedot{\futurelet\@let@token\@onedot}
\def\@onedot{\ifx\@let@token.\else.\null\fi\xspace}

\def\eg{\emph{e.g}\onedot} \def\Eg{\emph{E.g}\onedot}
\def\ie{\emph{i.e}\onedot} \def\Ie{\emph{I.e}\onedot}
\def\cf{\emph{c.f}\onedot} \def\Cf{\emph{C.f}\onedot}
\def\etc{\emph{etc}\onedot}
\def\vs{\emph{vs}\onedot}
\def\wrt{w.r.t\onedot} \def\dof{d.o.f\onedot}
\def\etal{\emph{et al}\onedot}
\makeatother

\iclrfinalcopy
\begin{document}
\fi

\title{StreamAV-Bench: A Comprehensive Benchmark for Streaming Audio-Video Generation}

\author{
Kaiqi Liu\textsuperscript{1,2} \quad
Haoxuan Zeng\textsuperscript{2} \quad
Jingqi Liu\textsuperscript{1,2} \quad
Jiacong Fang\textsuperscript{1,2} \quad
Ziqi Cai\textsuperscript{2} \\
Yunyao Mao\textsuperscript{3} \quad
Henglin Liu\textsuperscript{4} \quad
Yu Sheng\textsuperscript{5} \quad
Shuchen Weng\textsuperscript{1,2}\thanks{Corresponding authors.} \quad
Boxin Shi\textsuperscript{2*} \\
\textsuperscript{1}BAAI \quad
\textsuperscript{2}PKU \quad
\textsuperscript{3}Kling \quad
\textsuperscript{4}THU \quad
\textsuperscript{5}USTC \\[4pt]
{\normalfont\hypersetup{hidelinks}%
\href{https://liukqchoco.github.io/StreamAVBench/}{\textcolor[rgb]{0.21,0.49,0.74}{\texttt{https://liukqchoco.github.io/StreamAVBench/}}}}
}

\maketitle
\fancyhead{}

\begin{abstract}

Recent advancements in generative models are pushing video generation toward unbounded streaming audio-video generation for real-time interactive worlds. 
However, existing benchmarks primarily evaluate completed sequences and struggle to capture streaming properties. 
To bridge this gap, we introduce StreamAV-Bench, the first comprehensive benchmark tailored for streaming audio-video generation. 
StreamAV-Bench establishes a unified evaluation framework, including the progressive track for instruction adherence and long-horizon stability, and the interactive track for interactive response and state retention and reuse. With expert-verified evaluation cases across 32 fine-grained dimensions, we conduct an extensive evaluation of 13 representative systems. Our analysis reveals that current models suffer from temporal drift in progressive generation and responsiveness bottlenecks during interactive control. Based on a comprehensive failure analysis, we share insights to advance the development of native joint audio-video streaming models.

\end{abstract}

\begin{figure}[h]
\centering
\includegraphics[width=\linewidth]{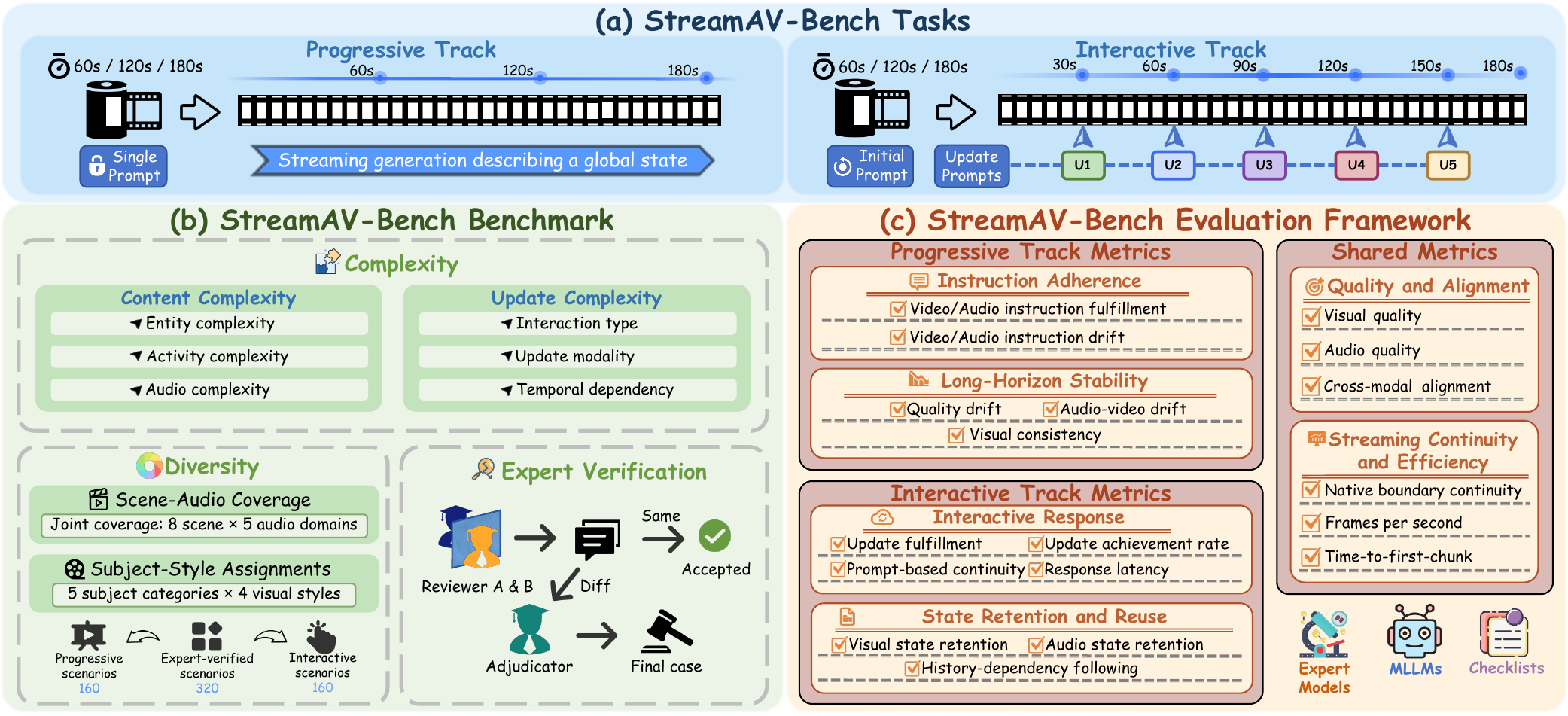}
\caption{We present StreamAV-Bench, a comprehensive benchmark for streaming
audio-video generation. (a) The benchmark includes a progressive track to
assess instruction adherence and long-horizon stability, and an interactive
track to measure interactive response alongside state retention and reuse.
(b) The benchmark covers content complexity in both tracks and update complexity
in the interactive track, with 320 expert-verified scenarios that span 8 scene
domains, 5 audio domains, 5 subject categories, and 4 visual styles. (c) The evaluation is
driven by a unified framework of 32 fine-grained dimensions, computed with
expert models, MLLMs, and corresponding checklists.}
\label{fig:overview}
\end{figure}

\section{Introduction}
Recent advances have pushed video generation beyond short clips toward long-horizon and interactive audio-video generation, paving the way for the real-time interactive world. Built upon the streaming paradigm, recent systems~\cite{su_2026_omniforcing,bai_2026_mainecoon,huang_2026_wanstreamer} generate these immersive worlds over unbounded durations.
In this setting, audio-video sequences are generated progressively over an expanding time horizon. Users can interactively provide free-form prompts, while previously generated clips are fixed as committed history.
Consequently, the model must maintain robust audio-video quality and cross-modal alignment, avoid temporal drift, execute dynamic switches in world states with rapid interactive response, and preserve boundary continuity as the generation progresses.
These requirements introduce novel failure modes that cannot be adequately captured by standard evaluation protocols.

Current benchmarks struggle to evaluate the properties of streaming audio-video generation. 
While largely exploring long-horizon generation, previous video-centric benchmarks~\cite{huang_2025_vbenchpp,zheng_2025_locot2v,yuan_2026_helios,liu_2026_iamflow,zhang_2026_mbench} lack audio-video quality assessment.
Meanwhile, existing audio-video generation benchmarks~\cite{hua_2025_vabench,cao_2025_t2avcompass,zhou_2026_avgenbench} focus primarily on fundamental metrics (\eg, audio-video quality, cross-modal alignment, and synchronization), with sparse attempts to explore long-horizon generation~\cite{liu_2026_longavcompass} and narratives with scripted prompts~\cite{wei_2026_msavbench}. 
Collectively, these benchmarks focus on completed audio-video outputs, neglecting the aforementioned streaming properties, leaving streaming audio-video generation insufficiently evaluated. 
We summarize the capabilities of existing benchmarks in \cref{tab:benchmark_comparison} to illustrate their specific focuses and their neglect of streaming properties.

In this paper, we introduce \textbf{StreamAV-Bench}, the first comprehensive benchmark for streaming audio-video generation. 
As shown in \cref{fig:overview}~(a), StreamAV-Bench includes a \textbf{progressive track} that uses a global prompt to evaluate progressively generated audio-video sequences for instruction adherence and long-horizon stability, and an \textbf{interactive track} that adopts an ordered prompt sequence to enable dynamic interaction and evaluate interactive response alongside state retention and reuse.
As shown in \cref{fig:overview}~(b), the benchmark considers two complementary sources of complexity: content complexity, characterized by entity, activity, and audio complexity in both tracks, and update complexity, characterized by interaction type, update modality, and temporal dependency in the interactive track. 
Collectively, the benchmark comprises 320 scenarios, spanning 8 scene domains, 5 audio domains, 5 subject categories, and 4 visual styles. These scenarios are evenly divided into the progressive and interactive tracks. To evaluate temporal scalability, we generate each scenario continuously for 180 seconds.
To ensure a rigorously curated benchmark, we further conduct expert verification on all cases to confirm prompt feasibility and eliminate semantic ambiguity.

Building on these two tracks and the resulting evaluation cases, StreamAV-Bench provides a unified evaluation framework with 32 fine-grained dimensions. As depicted in \cref{fig:overview}~(c), progressive metrics evaluate instruction adherence and long-horizon stability across continuous generation, while interactive metrics measure interactive response alongside state retention and reuse during dynamic updates. Additional shared metrics focus on streaming continuity and efficiency, as well as overall audio-video quality and alignment.

In summary, our main contributions are threefold:
\begin{itemize}
    \item We introduce StreamAV-Bench, the first comprehensive benchmark tailored for streaming audio-video generation, featuring distinct progressive and interactive tracks.
    \item We curate high-quality evaluation cases through strict independent expert verification and design a unified framework with 32 dimensions for systematic assessment.
    \item We conduct an extensive evaluation of 13 representative generation systems, spanning native streaming audio-video models and cascaded T2V and V2A pipelines.
\end{itemize}
We further reveal complementary strengths between the evaluated cascaded and native joint systems, dimension-specific long-horizon degradation, and a mismatch between update achievement and response latency. Additionally, we analyze modality-driven interaction bottlenecks, complemented by a qualitative failure analysis. Finally, we share insights to develop native joint audio-video streaming models with long-horizon stability and real-time interactive capabilities.

\begin{table*}[t]
\centering
\footnotesize
\setlength{\tabcolsep}{8pt}
\begin{adjustbox}{width=\textwidth,keepaspectratio}
\begin{tabular}{@{}l ccc cccccc@{}}
\toprule
\multirow[c]{2}{*}{\textbf{Benchmark}} &
\multicolumn{3}{c}{\textbf{Generation Setting}} &
\multicolumn{6}{c}{\textbf{Streaming Evaluation}} \\
\cmidrule(lr){2-4}\cmidrule(lr){5-10} &
\shortstack{\textbf{AV}\\\textbf{Quality}} &
\shortstack{\textbf{Long}\\\textbf{Horizon}} &
\shortstack{\textbf{Scripted}\\\textbf{Prompt}} &
\shortstack{\textbf{Temporal}\\\textbf{Drift}} &
\shortstack{\textbf{Dynamic}\\\textbf{Switch}} &
\shortstack{\textbf{Interactive}\\\textbf{Response}} &
\shortstack{\textbf{State Retention}\\\textbf{\& Reuse}} &
\shortstack{\textbf{Boundary}\\\textbf{Continuity}} &
\shortstack{\textbf{Streaming}\\\textbf{Efficiency}} 
 \\
\midrule
\multicolumn{10}{l}{\emph{Long-horizon video benchmarks}} \\
VBench-Long \cite{huang_2025_vbenchpp}
& \xmark & \cmark & \xmark & \xmark & \xmark & \xmark & \xmark & \xmark & \xmark \\
HeliosBench \cite{yuan_2026_helios}
& \xmark & \cmark & \xmark & \cmark & \xmark & \xmark & \xmark & \xmark & \cmark \\
NarrLV \cite{feng_2025_narrlv}
& \xmark & \cmark & \xmark & \xmark & \xmark & \xmark & \xmark & \xmark & \xmark \\
MPVBench \cite{cai_2024_ditctrl}
& \xmark & \cmark & \cmark & \xmark & \xmark & \xmark & \xmark & \cmark & \xmark \\
NarraStream-Bench \cite{liu_2026_iamflow}
& \xmark & \cmark & \cmark & \xmark & \cmark & \xmark & \cmark & \cmark & \cmark \\
MBench \cite{zhang_2026_mbench}
& \xmark & \cmark & \cmark & \xmark & \cmark & \cmark & \xmark & \xmark & \xmark \\
\midrule
\multicolumn{10}{l}{\emph{Audio-video generation benchmarks}} \\
VABench \cite{hua_2025_vabench}
& \cmark & \xmark & \xmark & \xmark & \xmark & \xmark & \xmark & \xmark & \xmark \\
T2AV-Compass \cite{cao_2025_t2avcompass}
& \cmark & \xmark & \xmark & \xmark & \xmark & \xmark & \xmark & \xmark & \xmark \\
AVGen-Bench \cite{zhou_2026_avgenbench}
& \cmark & \xmark & \xmark & \xmark & \xmark & \xmark & \xmark & \xmark & \xmark \\
LongAV-Compass \cite{liu_2026_longavcompass}
& \cmark & \cmark & \cmark & \xmark & \xmark & \xmark & \xmark & \cmark & \xmark \\
MSAVBench \cite{wei_2026_msavbench}
& \cmark & \xmark & \cmark & \xmark & \xmark & \xmark & \xmark & \xmark & \xmark \\
\midrule
\textbf{StreamAV-Bench (Ours)}
& \cmark & \cmark & \cmark & \cmark & \cmark & \cmark & \cmark & \cmark & \cmark \\
\bottomrule
\end{tabular}
\end{adjustbox}

\caption{Comparison with existing generative benchmarks. Our proposed StreamAV-Bench comprehensively evaluates the essential properties of streaming audio-video generation.}

\label{tab:benchmark_comparison}
\end{table*}

\section{Related Work}

\subsection{Streaming Audio-Video Generation}

To create immersive worlds, recent studies have increasingly focused on audio-video generation, either by developing cascaded pipelines that add audio to generated video~\cite{cheng_2024_mmaudio,shan_2025_hunyuanfoley} or by exploring joint audio-video generation models~\cite{ruan_2023_mmdiffusion,liu_2025_javisdit,low_2025_ovi,hacohen_2026_ltx2}. Extensive research has achieved streaming properties within the video modality. Specifically, to reduce accumulated errors during progressive generation, causal distillation and self-rollout training schedules~\cite{yin_2024_causvid,huang_2025_selfforcing} have been proposed. Advanced works~\cite{kodaira_2025_streamdit, yuan_2026_helios, liu_2025_rollingforcing, yang_2025_longlive} further introduce block-wise processing, rolling contexts, and bounded caches. 
Extending these concepts to audio-video generation, OmniForcing~\cite{su_2026_omniforcing} and Wan-Streamer~\cite{huang_2026_wanstreamer} adopt similar frameworks while maintaining audio-video synchronization, whereas MAVIN~\cite{liu_2026_mavin} and JoyAI-Echo~\cite{joyfuture_2026_joyaiecho} further explore controlled multi-shot and long-horizon audio-video generation. 
These advances support progressive and interactive streaming audio-video generation, highlighting the need for a comprehensive benchmark to evaluate their emerging capabilities.

\subsection{Video Generation Benchmarks}
Recent years have witnessed the emergence of video benchmarks to comprehensively evaluate generated video quality, temporal consistency, and prompt alignment~\cite{huang_2024_vbench}. 
Among them, FETV~\cite{liu_2023_fetv} and EvalCrafter~\cite{liu_2024_evalcrafter} provide human-aligned evaluations, while VBench++~\cite{huang_2025_vbenchpp} extends this coverage across various generation settings. 
Building upon these foundations, T2V-CompBench~\cite{sun_2025_t2vcompbench}, DEVIL~\cite{liao_2024_devil}, and TC-Bench~\cite{feng_2025_tcbench} target compositional instruction following, video dynamics, and temporal state transitions.
For long-video generation, LV-Bench~\cite{zhang_2025_bife} uses minute-long videos with chunk-level annotations to quantify visual degradation, whereas LoCoT2V-Bench~\cite{zheng_2025_locot2v} evaluates event-level alignment and narrative realization under complex prompts. 
Under multi-prompt settings, NarraStream-Bench~\cite{liu_2026_iamflow} evaluates instruction following across generated segments, and MBench~\cite{zhang_2026_mbench} focuses on long-range memory through entity, environment, and causal consistency. 
Despite effectively exposing the limitations of video generation models, these works remain video-centric, overlooking the accompanying audio modality.

\subsection{Audio-Video Generation Benchmarks}
Recent audio-video benchmarks~\cite{mao_2024_tavgbench,hua_2025_vabench,cao_2025_t2avcompass,zhou_2026_avgenbench} have constructed comprehensive evaluation protocols encompassing video and audio quality, cross-modal semantics, temporal synchronization, instruction following, and perceptual realism. Targeting specific settings, MTAVG-Bench~\cite{zhou_2026_mtavg} and FoleyBench~\cite{dixit_2025_foleybench} examine visually grounded sound and multi-talker scenarios. Broadening this scope, MSAVBench~\cite{wei_2026_msavbench} extends to multi-shot generation through global, cross-shot, and intra-shot assessments, while LongAV-Compass~\cite{liu_2026_longavcompass} investigates minute-long generation under diverse conditioning modalities. Although these benchmarks systematically cover short-form, multi-shot, and minute-long audio-video outputs, they are fundamentally designed for static and completed generations. Consequently, they cannot be directly applied to streaming systems for real-time interactive worlds. 

\section{StreamAV-Bench}
\subsection{Task Definition}
Streaming audio-video generation models produce synchronized audio-video sequences incrementally over an expanding time horizon. Previously generated sequences become committed history and condition subsequent generation. The prompt may remain fixed or be updated interactively during generation, corresponding to progressive and interactive streaming, respectively. For the created audio-video world, we define its state as the audio-video content established by a specific prompt, which is expected to remain valid unless modified by a subsequent prompt update and may be reused by future instructions.

To provide a comprehensive evaluation, StreamAV-Bench introduces a unified benchmarking framework to evaluate streaming models across two tracks: The progressive track evaluates models' instruction adherence and long-horizon generation stability using a 180-second audio-video sequence generated from a single prompt describing a global state. The interactive track instead evaluates the model's interactive response alongside state retention and reuse using an ordered prompt sequence, where the initial prompt defines the starting scenario, and five subsequent prompts are provided every 30 seconds to update visual content, audio content, or both. For both tracks, we additionally evaluate the first 60, 120, and 180 seconds of the same rollout. This allows us to examine how performance changes with generation length in both the progressive and interactive tracks.

\begin{figure*}[t]
\centering
\includegraphics[width=\textwidth]{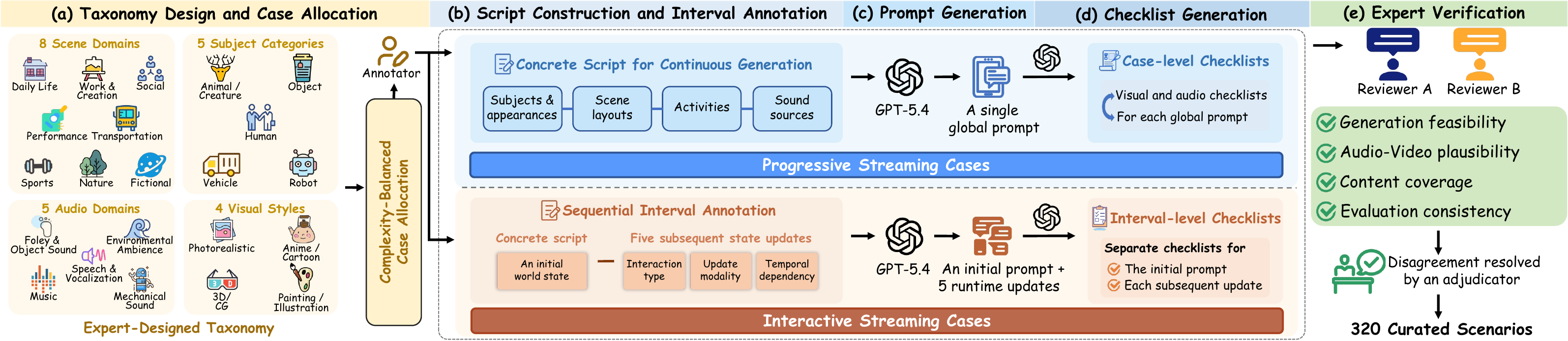}

\caption{Overview of the data construction pipeline for StreamAV-Bench. 
(a) Domain experts define scene and audio domains, subject categories, and visual styles, and allocate cases for the two tracks, ensuring complexity balance. 
(b) Annotators script each case with annotations: the progressive track defines the entire sequence, while the interactive track specifies an interval sequence for dynamic state updates. 
(c) GPT-5.4 converts these annotations into a single global prompt per case for the progressive track, and an initial prompt followed by five runtime updates for the interactive track. 
(d) These prompts are used to derive verifiable checklists, evaluating case-level instruction adherence and stability, and interval-level interactive response, state retention, and reuse.
(e) Two experts independently review the prompts and checklists, with disagreements adjudicated by a third expert, producing 320 curated scenarios.}

\label{fig:construction}
\end{figure*}

\subsection{Benchmark Construction}

To ensure a diverse and reliable benchmark for streaming audio-video generation, we employ a five-stage construction pipeline, as illustrated in~\cref{fig:construction}.

\noindent\textbf{Taxonomy design and case allocation.}
Domain experts first define a taxonomy including 8 scene domains, 5 audio domains, 5 subject categories, and 4 visual styles. To ensure balanced domain coverage, we construct 160 scenario themes, each used once in the progressive track and once in the interactive track, yielding 320 scenarios. For both tracks, three content complexity dimensions (entity, activity, and audio complexity) define the base content. For each interactive scenario, we additionally specify how its five runtime updates are distributed across three update complexity dimensions: interaction type, update modality, and temporal dependency.

\noindent\textbf{Script construction and interval annotation.}
For the allocated scenarios, annotators develop concrete scripts to describe subjects and appearances, scene layouts, activities, and sound sources. Using this context, they produce structured annotations to organize the script over time. For the progressive track, these annotations bind the expected content across the entire sequence. For the interactive track, they define an initial world state and subsequent state updates for each following 30-second interval.

\noindent\textbf{Prompt generation.}
Based on the annotated scripts, we employ an LLM~\cite{openai_2026_gpt54} to generate text prompts. Specifically, the progressive track assigns a single prompt to describe the global content (\eg, scene, visual style, subjects, ongoing activities, and audio). In contrast, the interactive track utilizes an initial prompt to establish the initial world, followed by five textual updates for subsequent intervals, each specifying the expected targets and state changes.

\noindent\textbf{Checklist generation.}
Following prompt generation, we utilize an LLM~\cite{openai_2026_gpt54} to derive verifiable checklists, decomposing explicit audio-video requirements into measurable criteria. Progressive cases associate visual and audio checklists with each prompt for case-level instruction adherence and stability, while interactive cases require separate criteria for the initial prompt and each subsequent update for interval-level evaluation.

\noindent\textbf{Expert verification.}
We assign two domain experts as independent reviewers and a third as an adjudicator. The reviewers examine the prompts, checklists, and update targets against the corresponding scripts and interval annotations. They assess generation feasibility, audio-video plausibility, content coverage, and evaluation consistency. If a generated prompt omits required script content, a checklist or update target is not supported by its prompt, or an update conflicts with an established state, the case is revised and reviewed again. The adjudicator resolves any disagreements, producing a final curated set of 320 scenarios.

\subsection{Benchmark Statistics}

\begin{figure*}[t]
\centering
\includegraphics[width=\textwidth]{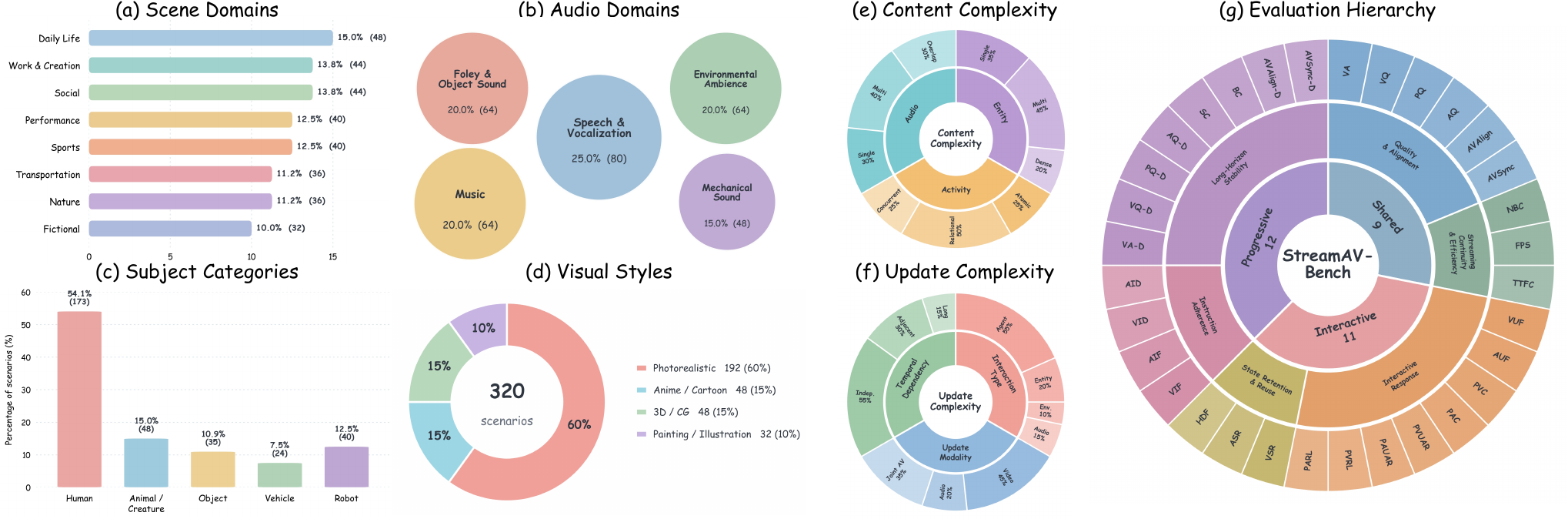}

\caption{Detailed data statistics and the evaluation framework of StreamAV-Bench.
(a)--(d) Distributions across 8 scene domains, 5 audio domains, 5 subject categories, and 4 visual styles.
(e)--(f) Content and update complexity factors, respectively.
(g) Overview of the evaluation framework, categorizing 32 fine-grained evaluation dimensions.}
\label{fig:distribution}

\end{figure*}

We detail the statistics of StreamAV-Bench to illustrate its structural distribution and complexity, as shown in \cref{fig:distribution}.

\noindent \textbf{Content distribution.}
StreamAV-Bench exhibits comprehensive scene-audio coverage and diverse subject and visual-style distributions. As shown in~\cref{fig:distribution}~(a)--(b), it contains 320 scenarios spanning 8 scene and 5 audio domains, with 32--48 scenarios per scene domain and 48--80 per audio domain. As detailed in~\cref{fig:distribution}~(c)--(d), its annotations encompass 5 subject categories and 4 visual styles, including 192 photorealistic and 128 non-photorealistic scenarios. These orthogonal combinations support robust evaluation across diverse audio-video settings.

\noindent\textbf{Content and update complexity.}
StreamAV-Bench defines specific complexity factors for both the progressive and interactive tracks, covering their diverse combinations to comprehensively evaluate streaming audio-video generation.
As shown in~\cref{fig:distribution}~(e), both tracks evaluate content complexity across three specific dimensions.
\textit{Entity complexity} reveals the visual state complexity (\eg, multiple subjects).
\textit{Activity complexity} represents the activity structure (\eg, concurrent activities).
\textit{Audio complexity} captures the sound-source complexity (\eg, overlapping sounds).
Together, these dimensions determine the overall complexity of long-horizon audio-video generation.
As illustrated in~\cref{fig:distribution}~(f), the interactive track captures update complexity along three axes.
\textit{Interaction type} reveals the requested operation complexity (\eg, agent action).
\textit{Update modality} indicates the modality editing complexity (\eg, video, audio, or both).
\textit{Temporal dependency} captures the history-dependency complexity (\eg, a long-range dependency on an earlier state).
Collectively, these dimensions comprehensively determine the complexity for runtime state manipulation.

\subsection{Evaluation Framework}
\label{sec:evaluation}
Our evaluation framework combines tailored expert models with Multimodal Large Language Model (MLLM) assessments guided by case-specific checklists.
As shown in~\cref{fig:distribution}~(g), our evaluation metrics comprise 32 fine-grained dimensions organized into six categories.
Specifically, the progressive track primarily measures instruction adherence and long-horizon stability, while the interactive track focuses on interactive response and state retention and reuse.
For both tracks, we additionally introduce shared metrics to evaluate streaming continuity and efficiency, as well as quality and alignment.

\noindent\textbf{Quality and alignment.}
We assess generated audio-video content along three complementary dimensions: visual quality, audio quality, and cross-modal alignment. For visual quality, visual aesthetics (\textit{VA}) uses the LAION Aesthetic Predictor~\cite{laion_2022_aesthetic} to measure visual appeal, while visual quality (\textit{VQ}) employs Gemini 3.1 Pro~\cite{google_2026_gemini31pro} to assess visual fidelity, subject integrity, motion naturalness, and visual artifacts. For audio quality, production quality (\textit{PQ}) uses Audiobox Aesthetics~\cite{tjandra_2025_audiobox} to measure perceptual and production fidelity, while audio quality (\textit{AQ}) employs Gemini 3.1 Pro to assess audio naturalness and artifacts. For cross-modal alignment, \textit{AVAlign} uses ImageBind similarity~\cite{girdhar_2023_imagebind} to measure semantic correspondence, whereas \textit{AVSync} uses Synchformer~\cite{iashin_2024_synchformer} to estimate temporal synchronization.

\noindent\textbf{Instruction adherence.}
For each progressive case, we use Gemini 3.1 Pro with fixed visual and audio checklists derived from the global prompt. Video instruction fulfillment (\textit{VIF}) and audio instruction fulfillment (\textit{AIF}) assess how completely the corresponding requirements are realized on a 1--5 scale. Extending the endpoint comparison used by HeliosBench~\cite{yuan_2026_helios}, video instruction drift (\textit{VID}) and audio instruction drift (\textit{AID}) reuse the same checklists to measure the change in adherence between the first and last 30-second intervals.

\noindent\textbf{Long-horizon stability.}
To evaluate long-horizon stability, the progressive track examines quality drift, visual consistency, and audio-video drift. Adapting the endpoint comparison used by HeliosBench~\cite{yuan_2026_helios}, we compute the absolute changes in visual aesthetics (\textit{VA-D}), visual quality (\textit{VQ-D}), production quality (\textit{PQ-D}), and audio quality (\textit{AQ-D}) between the first and last 30-second intervals. Subject consistency (\textit{SC}) and background consistency (\textit{BC}) follow VBench-Long~\cite{huang_2025_vbenchpp} to assess the temporal consistency of subject appearance and background scenes, respectively. Finally, \textit{AVAlign-D} and \textit{AVSync-D} measure endpoint changes in audio-video semantic correspondence and temporal synchronization.

\noindent\textbf{Interactive response.}
The interactive track evaluates the model's capacity for dynamic state updates. Video update fulfillment (\textit{VUF}) and audio update fulfillment (\textit{AUF}) use Gemini 3.1 Pro to assess how completely each update's visual and audio targets are realized. During prompt updates, we calculate prompt-based visual and audio continuity (\textit{PVC}, \textit{PAC}) to further assess temporal transition smoothness. Finally, prompt-based visual and audio update achievement rates (\textit{PVUAR} and \textit{PAUAR}) measure the proportion of updates whose core targets are realized, while the corresponding response latencies (\textit{PVRL} and \textit{PARL}) measure how quickly successful realization occurs.

\noindent\textbf{State retention and reuse.}
To evaluate whether required elements (\eg, subject appearance and ongoing audio) are consistently preserved, visual state retention (\textit{VSR}) and audio state retention (\textit{ASR}) are introduced. \textit{VSR} and \textit{ASR} compare the intervals before and after each update while accounting for the changes requested by that update. To further evaluate state reuse, history-dependency following (\textit{HDF}) measures whether the current instruction is correctly executed using relevant content from earlier interactions. All three metrics use Gemini 3.1 Pro on 1--5 scales.

\noindent\textbf{Streaming continuity and efficiency.}
At the boundaries of consecutive model-native generation units, native boundary continuity (\textit{NBC}) assesses technical continuity by detecting boundary artifacts such as black frames, repeated frames, freezing, and flashes. We report generation speed in frames per second (\textit{FPS}) under each system's default inference configuration, and quantify startup latency as time-to-first-chunk (\textit{TTFC}).

\section{Experiments}

\subsection{Experimental Setup}

\noindent\textbf{Evaluated models.}
We comprehensively evaluate 13 systems across 2 distinct generation paradigms.
Specifically, PixVerse R1~\cite{pixverse_2026_r1}, HappyOyster~\cite{alibaba_2026_happyoyster}, and OmniForcing~\cite{su_2026_omniforcing} represent native joint streaming audio-video models.
To establish cascaded streaming pipelines, we pair a dedicated audio generator (HunyuanVideo-Foley~\cite{shan_2025_hunyuanfoley}) with various native streaming video models: Odyssey-2~\cite{odyssey_2025_odyssey2}, Self-Forcing~\cite{huang_2025_selfforcing}, LongLive~\cite{yang_2025_longlive}, Rolling-Forcing~\cite{liu_2025_rollingforcing}, Deep Forcing~\cite{yi_2025_deepforcing}, MemFlow~\cite{ji_2025_memflow}, Causal-Forcing~\cite{zhu_2026_causalforcing}, Helios~\cite{yuan_2026_helios}, SWIFT~\cite{tan_2026_swift}, and IAMFlow~\cite{liu_2026_iamflow}.

\noindent\textbf{Implementation details.}
We evaluate open-source models using their official checkpoints and inference pipelines on NVIDIA H800 GPUs, while commercial systems are accessed via their official web interfaces. All models are deployed with their default recommended generation configurations. To ensure a consistent evaluation protocol, all systems process identical benchmark inputs: a single global prompt for the progressive track, and an ordered prompt sequence with fixed 30-second intervals for the interactive track.

\noindent\textbf{Evaluation protocol.}
All systems are evaluated using the framework detailed in \cref{sec:evaluation}, utilizing case-specific checklists for instruction-conditioned metrics.
Crucially, we strictly align our evaluation metrics with the established track formulations.
Shared metrics for streaming continuity, efficiency, quality, and alignment are computed across both tracks.
Instruction adherence and long-horizon stability are evaluated on the progressive track, while interactive response and state retention and reuse are assessed on the interactive track.
During statistical aggregation, observations are first compiled within each individual case and subsequently averaged uniformly across cases.

\begin{table*}[!t]
\centering
\scriptsize
\setlength{\tabcolsep}{3pt}
\renewcommand{\arraystretch}{0.88}

\begin{tabular*}{\textwidth}{@{\extracolsep{\fill}}l*{9}{c}@{}}
\toprule
\multirow[c]{2}{*}{\textbf{Method}} &
\multicolumn{6}{c}{\textbf{Quality and Alignment}} &
\multicolumn{3}{c}{\textbf{Streaming Continuity and Efficiency}} \\
\cmidrule(lr){2-7}\cmidrule(lr){8-10}
& \textbf{VA}$\uparrow$ & \textbf{VQ}$\uparrow$
& \textbf{PQ}$\uparrow$ & \textbf{AQ}$\uparrow$
& \textbf{AVAlign}$\uparrow$ & \textbf{AVSync}$\downarrow$
& \textbf{NBC}$\uparrow$ & \textbf{FPS}$\uparrow$ & \textbf{TTFC}$\downarrow$ \\
\midrule
\multicolumn{10}{l}{\emph{Native joint streaming audio-video models}} \\
PixVerse R1 & 0.529 & 2.428 & 6.348 & \underline{2.810} & 0.234 & \textbf{0.855} & -- & -- & -- \\
HappyOyster & 0.529 & 2.785 & \textbf{6.817} & \textbf{3.157} & 0.206 & 1.044 & -- & -- & -- \\
OmniForcing & 0.554 & 2.374 & 6.351 & 2.528 & 0.119 & 1.423 & \underline{98.808} & 12.799 & 3.998 \\
\midrule
\multicolumn{10}{l}{\emph{Cascaded streaming audio-video pipelines}} \\
Odyssey-2 & 0.531 & 2.820 & 6.426 & 2.659 & 0.260 & 0.946 & -- & -- & -- \\
Self-Forcing & 0.585 & 2.753 & 6.453 & 2.623 & 0.256 & 0.919 & 94.359 & 15.628 & 1.278 \\
LongLive & \underline{0.605} & 2.768 & 6.459 & 2.696 & 0.272 & 0.969 & 85.418 & 10.125 & 1.381 \\
Rolling-Forcing & 0.572 & 2.757 & 6.393 & 2.673 & 0.265 & 1.078 & 96.657 & 10.603 & 2.962 \\
Deep Forcing & 0.597 & 2.632 & \underline{6.523} & 2.668 & \textbf{0.280} & \underline{0.878} & 92.482 & 11.065 & 1.397 \\
MemFlow & 0.600 & \underline{2.836} & 6.407 & 2.651 & 0.269 & 1.035 & 87.724 & 10.840 & 1.343 \\
Causal-Forcing & 0.556 & 2.397 & 6.383 & 2.714 & \underline{0.280} & 1.093 & \textbf{99.558} & 15.421 & 1.284 \\
Helios & 0.470 & 2.176 & 6.241 & 2.521 & 0.200 & 0.927 & 83.657 & 10.231 & 8.059 \\
SWIFT & \textbf{0.605} & 2.735 & 6.483 & 2.680 & 0.270 & 0.970 & 87.043 & 12.359 & 1.496 \\
IAMFlow & 0.602 & \textbf{2.840} & 6.415 & 2.625 & 0.268 & 1.016 & 86.957 & 6.049 & 1.772 \\
\bottomrule
\end{tabular*}

\vspace{3pt}

\begin{tabular*}{\textwidth}{@{\extracolsep{\fill}}l*{12}{c}@{}}
\toprule
\multirow[c]{2}{*}{\textbf{Method}} &
\multicolumn{4}{c}{\textbf{Instruction Adherence}} &
\multicolumn{8}{c}{\textbf{Long-Horizon Stability}} \\
\cmidrule(lr){2-5}\cmidrule(lr){6-13}
& \textbf{VIF}$\uparrow$ & \textbf{AIF}$\uparrow$
& \textbf{VID}$\downarrow$ & \textbf{AID}$\downarrow$
& \textbf{VA-D}$\downarrow$ & \textbf{VQ-D}$\downarrow$ & \textbf{PQ-D}$\downarrow$
& \textbf{AQ-D}$\downarrow$ & \textbf{SC}$\uparrow$
& \textbf{BC}$\uparrow$ & \textbf{AVAlign-D}$\downarrow$
& \textbf{AVSync-D}$\downarrow$ \\
\midrule
\multicolumn{13}{l}{\emph{Native joint streaming audio-video models}} \\
PixVerse R1 & 2.976 & 2.511 & 0.484 & \underline{0.425} & 0.032 & \underline{0.253} & 0.421 & 0.626 & 0.907 & 0.930 & 0.065 & 0.579 \\
HappyOyster & \textbf{3.565} & 2.692 & 0.510 & 0.508 & 0.032 & 0.417 & 0.623 & \textbf{0.411} & 0.904 & 0.934 & 0.077 & 0.534 \\
OmniForcing & 2.423 & 2.343 & 0.572 & \textbf{0.277} & 0.083 & 0.273 & 0.429 & 0.506 & 0.962 & 0.953 & 0.081 & 0.479 \\
\midrule
\multicolumn{13}{l}{\emph{Cascaded streaming audio-video pipelines}} \\
Odyssey-2 & 3.106 & 2.778 & 0.500 & 0.566 & 0.040 & 0.463 & 0.362 & 0.456 & 0.976 & 0.966 & 0.064 & 0.548 \\
Self-Forcing & 3.019 & 2.810 & 0.373 & 0.457 & 0.025 & 0.323 & 0.341 & 0.488 & 0.981 & 0.969 & 0.054 & \underline{0.476} \\
LongLive & 3.164 & 2.814 & 0.317 & 0.581 & 0.023 & 0.367 & \underline{0.280} & 0.475 & 0.986 & 0.973 & \underline{0.052} & 0.476 \\
Rolling-Forcing & 3.009 & \underline{2.858} & 0.425 & 0.474 & 0.043 & 0.427 & 0.329 & \underline{0.425} & 0.980 & 0.966 & 0.064 & 0.506 \\
Deep Forcing & 3.041 & 2.629 & 0.359 & 0.545 & 0.023 & 0.334 & 0.289 & 0.497 & 0.979 & 0.967 & 0.062 & 0.553 \\
MemFlow & 3.142 & 2.854 & 0.333 & 0.492 & 0.028 & 0.502 & 0.319 & 0.503 & \textbf{0.986} & 0.972 & 0.054 & \textbf{0.474} \\
Causal-Forcing & 2.834 & 2.729 & 0.364 & 0.615 & 0.030 & \textbf{0.189} & 0.320 & 0.509 & 0.952 & 0.946 & 0.058 & 0.491 \\
Helios & 2.977 & 2.712 & 0.694 & 0.566 & 0.149 & 0.306 & 0.649 & 0.513 & 0.967 & 0.967 & 0.115 & 0.514 \\
SWIFT & 3.152 & \textbf{2.887} & \underline{0.305} & 0.515 & \textbf{0.022} & 0.411 & \textbf{0.272} & 0.463 & 0.986 & \underline{0.973} & \textbf{0.049} & 0.501 \\
IAMFlow & \underline{3.192} & 2.789 & \textbf{0.302} & 0.600 & \underline{0.023} & 0.425 & 0.331 & 0.563 & \underline{0.986} & \textbf{0.973} & 0.057 & 0.490 \\
\bottomrule
\end{tabular*}

\vspace{3pt}

\begin{tabular*}{\textwidth}{@{\extracolsep{\fill}}l*{11}{c}@{}}
\toprule
\multirow[c]{2}{*}{\textbf{Method}} &
\multicolumn{8}{c}{\textbf{Interactive Response}} &
\multicolumn{3}{c}{\makebox[0pt]{\textbf{State Retention and Reuse}}} \\
\cmidrule(lr){2-9}\cmidrule(lr){10-12}
& \textbf{VUF}$\uparrow$ & \textbf{AUF}$\uparrow$
& \textbf{PVC}$\uparrow$ & \textbf{PAC}$\uparrow$
& \textbf{PVUAR}$\uparrow$ & \textbf{PAUAR}$\uparrow$
& \textbf{PVRL}$\downarrow$ & \textbf{PARL}$\downarrow$
& \textbf{VSR}$\uparrow$ & \textbf{ASR}$\uparrow$
& \textbf{HDF}$\uparrow$ \\
\midrule
\multicolumn{12}{l}{\emph{Native joint streaming audio-video models}} \\
PixVerse R1 & \underline{2.573} & \textbf{2.895} & \textbf{4.278} & 2.544 & \underline{0.350} & \underline{0.466} & 12.992 & \underline{10.352} & 2.430 & \textbf{4.108} & 2.461 \\
HappyOyster & \textbf{2.975} & \underline{2.879} & 3.744 & \textbf{3.509} & \textbf{0.518} & \textbf{0.488} & 14.484 & 13.621 & 2.799 & \underline{3.772} & \textbf{2.817} \\
OmniForcing & 1.848 & 2.242 & 4.161 & \underline{3.369} & 0.053 & 0.316 & 13.295 & \textbf{7.925} & 2.624 & 3.388 & 2.487 \\
\midrule
\multicolumn{12}{l}{\emph{Cascaded streaming audio-video pipelines}} \\
Odyssey-2 & 2.408 & 2.244 & 3.919 & 1.749 & 0.274 & 0.308 & 18.758 & 15.733 & 2.836 & 2.707 & \underline{2.554} \\
Self-Forcing & 2.237 & 2.339 & 3.912 & 1.784 & 0.185 & 0.334 & 14.164 & 12.369 & \underline{2.851} & 2.813 & 2.348 \\
LongLive & 2.341 & 2.138 & 3.631 & 1.809 & 0.241 & 0.266 & 11.792 & 13.078 & 2.724 & 2.712 & 2.212 \\
Rolling-Forcing & 2.315 & 2.221 & 4.086 & 1.800 & 0.217 & 0.298 & 13.234 & 12.820 & 2.542 & 2.597 & 2.502 \\
Deep Forcing & 2.295 & 2.146 & 3.879 & 1.770 & 0.220 & 0.288 & 13.433 & 12.650 & 2.449 & 2.601 & 2.206 \\
MemFlow & 2.309 & 2.286 & 3.630 & 1.880 & 0.217 & 0.326 & 13.698 & 13.154 & \textbf{2.961} & 2.879 & 2.385 \\
Causal-Forcing & 2.284 & 2.165 & \underline{4.218} & 1.715 & 0.219 & 0.283 & 14.032 & 12.680 & 2.218 & 2.468 & 2.357 \\
Helios & 2.253 & 2.461 & 3.317 & 1.644 & 0.197 & 0.354 & 14.043 & 12.207 & 2.405 & 1.985 & 2.478 \\
SWIFT & 2.357 & 2.102 & 3.615 & 1.856 & 0.246 & 0.276 & \textbf{10.340} & 12.094 & 2.609 & 2.705 & 2.322 \\
IAMFlow & 2.319 & 2.190 & 3.655 & 1.819 & 0.222 & 0.297 & \underline{11.683} & 13.001 & 2.836 & 2.903 & 2.424 \\
\bottomrule
\end{tabular*}

\caption{Evaluation results on StreamAV-Bench. Metrics for quality, alignment, streaming continuity, and efficiency are evaluated on both tracks. Instruction adherence and long-horizon stability are evaluated on the progressive track, while interactive response and state retention and reuse are evaluated on the interactive track. The best and second-best results are highlighted in \textbf{bold} and \underline{underlined}. Dashes~(--) denote unavailable results.}

\label{tab:main_results}\end{table*}

\subsection{Main Results}

Table~\ref{tab:main_results} summarizes the overall performance and highlights three findings about current streaming generation systems.

\noindent\textbf{Performance is distributed across paradigms.}
Among the evaluated systems, native joint streaming audio-video models achieve the strongest audio fidelity and synchronization (\textit{PQ}, \textit{AQ}, and \textit{AVSync}) and lead nine of the eleven interactive-track metrics. Cascaded pipelines based on native streaming T2V instead lead in visual quality and semantic alignment (\textit{VA}, \textit{VQ}, and \textit{AVAlign}), seven of the eight long-horizon stability metrics, and streaming continuity (\textit{NBC}). Instruction adherence remains mixed: native joint systems lead in video instruction fulfillment (\textit{VIF}) and audio instruction drift (\textit{AID}), whereas cascaded pipelines lead in audio instruction fulfillment (\textit{AIF}) and video instruction drift (\textit{VID}). Overall, no single system dominates the full evaluation suite.

\noindent\textbf{High absolute performance does not ensure temporal stability.}
The results show that high generation quality or instruction fulfillment does not necessarily imply low temporal drift. IAMFlow achieves the highest visual quality (\textit{VQ}) score of 2.840 but a visual quality drift (\textit{VQ-D}) of 0.425, whereas Causal-Forcing records the lowest \textit{VQ-D} of 0.189 despite a substantially lower \textit{VQ} score of 2.397. A similar pattern appears in audio instruction adherence: SWIFT achieves the highest audio instruction fulfillment (\textit{AIF}) score of 2.887 but an audio instruction drift (\textit{AID}) of 0.515, while OmniForcing records the lowest \textit{AID} of 0.277 despite the lowest \textit{AIF} score of 2.343. These results show that low drift may coexist with lower absolute performance, underscoring the need to evaluate generation quality and instruction fulfillment jointly with their temporal changes.

\noindent\textbf{Fast response does not guarantee successful interaction.}
The interactive-track results show that update achievement and response latency do not necessarily align. HappyOyster obtains the highest prompt-based visual and audio update achievement rates (\textit{PVUAR} and \textit{PAUAR}) of 0.518 and 0.488, respectively, but its corresponding response latencies (\textit{PVRL} and \textit{PARL}) are 14.484 and 13.621 seconds. In contrast, SWIFT achieves the lowest prompt-based visual response latency (\textit{PVRL}) of 10.340 seconds but a \textit{PVUAR} of only 0.246, while OmniForcing achieves the lowest prompt-based audio response latency (\textit{PARL}) of 7.925 seconds but a \textit{PAUAR} of 0.316. These results show that low latency among achieved updates does not imply reliable update achievement; achievement rates and response latencies must therefore be interpreted jointly. Meanwhile, MemFlow records the highest visual state retention (\textit{VSR}) score of 2.961, while no system exceeds 2.817 in history-dependency following (\textit{HDF}), underscoring the difficulty of retaining and reusing state during interaction.

\subsection{Analysis and Findings}

\begin{figure*}[t]
\centering
\includegraphics[width=\textwidth]{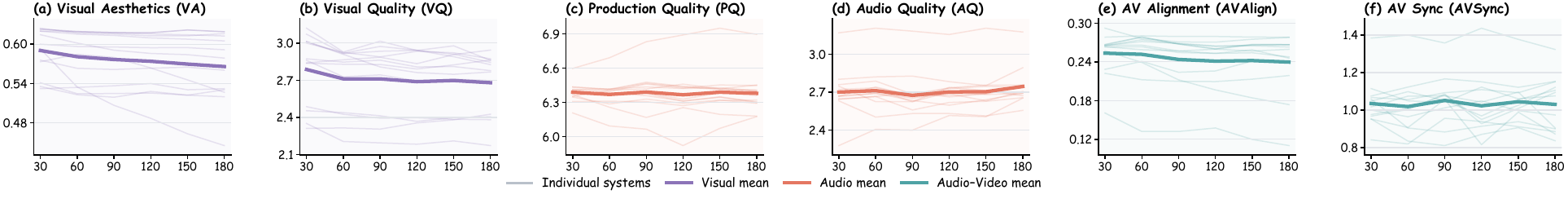}
\caption{Temporal evolution of progressive-track quality and audio-video alignment over 180 seconds. Thin curves show individual systems, while thick curves show the mean across 13 systems. Higher is better for VA, VQ, PQ, AQ, and AVAlign, whereas lower is better for AVSync.}
\label{fig:temporal_evolution}
\end{figure*}

\noindent\textbf{Dimension-specific quality evolution.}
We track 6 quality and alignment dimensions across the progressive track. Over 180 seconds, degradation slope captures the overall direction across all segments, while trajectory variation measures adjacent-segment fluctuations. As shown in \cref{fig:temporal_evolution}, visual aesthetics (\textit{VA}), audio-video alignment (\textit{AVAlign}), and visual quality (\textit{VQ}) degrade in 13, 12, and 10 of the 13 systems, respectively. By contrast, production quality (\textit{PQ}), audio quality (\textit{AQ}), and audio-video synchronization (\textit{AVSync}) exhibit mixed directions, degrading in 9, 6, and 7 systems. Trajectory variation further reveals instability concealed by the overall trend: Odyssey-2 and PixVerse R1 show improving \textit{VQ} and \textit{AQ} trends ($-0.056$ and $-0.019$), yet the largest variations in their respective dimensions (0.423 and 0.508). Meanwhile, SWIFT exhibits both the steepest \textit{AVSync} degradation (0.062) and the largest \textit{AVSync} variation (0.536). Consequently, streaming evolution is dimension-specific: visual quality and semantic alignment degrade most consistently, whereas audio quality and temporal synchronization often fluctuate rather than decline monotonically.

\noindent\textbf{Generation length exposes track-specific bottlenecks.}
We compare performance at 60, 120, and 180 seconds across both tracks. Averaged across the 13 systems, the progressive track exhibits increasing quality drift: visual aesthetics drift (\textit{VA-D}) rises from 0.025 to 0.036 and 0.042, while visual quality drift (\textit{VQ-D}) increases from 0.313 to 0.347 and 0.361. Production and audio quality drift (\textit{PQ-D} and \textit{AQ-D}) follow the same pattern, whereas audio-video alignment and synchronization drift (\textit{AVAlign-D} and \textit{AVSync-D}) change little beyond 120 seconds. In the interactive track, visual update fulfillment (\textit{VUF}) and achievement (\textit{PVUAR}) remain stable, whereas audio update fulfillment (\textit{AUF}) decreases from 2.481 to 2.340 and 2.332, and audio update achievement (\textit{PAUAR}) decreases from 0.361 to 0.333 and 0.331. Visual and audio response latencies (\textit{PVRL} and \textit{PARL}) both increase by approximately 1.3 seconds. Audio state retention (\textit{ASR}) also declines from 3.022 to 2.952 and 2.895, while visual state retention (\textit{VSR}) remains stable. Consequently, longer rollouts accumulate quality drift in progressive generation while primarily exposing audio-side degradation and slower responses under repeated interaction.

\noindent\textbf{Content complexity exposes a gap between content realization and temporal preservation.}
We analyze the progressive track across entity, activity, and audio complexity. Entity density produces the clearest visual degradation: visual instruction fulfillment (\textit{VIF}) decreases from 3.090 for single-entity cases to 2.816 for dense cases, while visual quality (\textit{VQ}) decreases monotonically from 2.760 to 2.651. Activity complexity reveals a similar visual bottleneck, with \textit{VQ} decreasing by 0.129 from atomic to concurrent activities without a corresponding decline in audio instruction fulfillment (\textit{AIF}). Multi-source cases attain the lowest \textit{AIF} (2.523), highlighting the challenge of satisfying multiple sound requirements. In contrast, subject consistency (\textit{SC}) and background consistency (\textit{BC}) remain nearly unchanged across all three factors. Consequently, increasing content complexity primarily disrupts the realization of required visual content, while subject appearance and background scenes remain comparatively stable.

\noindent\textbf{Visual state changes and history reuse remain interaction bottlenecks.}
For the interactive track, visual and audio update fulfillment (\textit{VUF} and \textit{AUF}) are comparable (2.347 vs. 2.332). However, visual updates have lower achievement rates than audio updates (\textit{PVUAR}/\textit{PAUAR}: 0.243/0.331) and respond more slowly (\textit{PVRL}/\textit{PARL}: 13.534/12.437 seconds). Direct agent actions also attain higher \textit{VUF} (2.418) than entity-state and environment changes (2.274 and 2.283). For history-dependent updates, \textit{HDF} reaches only 2.477 for adjacent and 2.321 for long-range dependencies. Therefore, the primary interaction bottlenecks lie in reliably realizing visual changes, revising established states, and reusing earlier information.

\begin{figure*}[t]
\centering
\includegraphics[width=\textwidth]{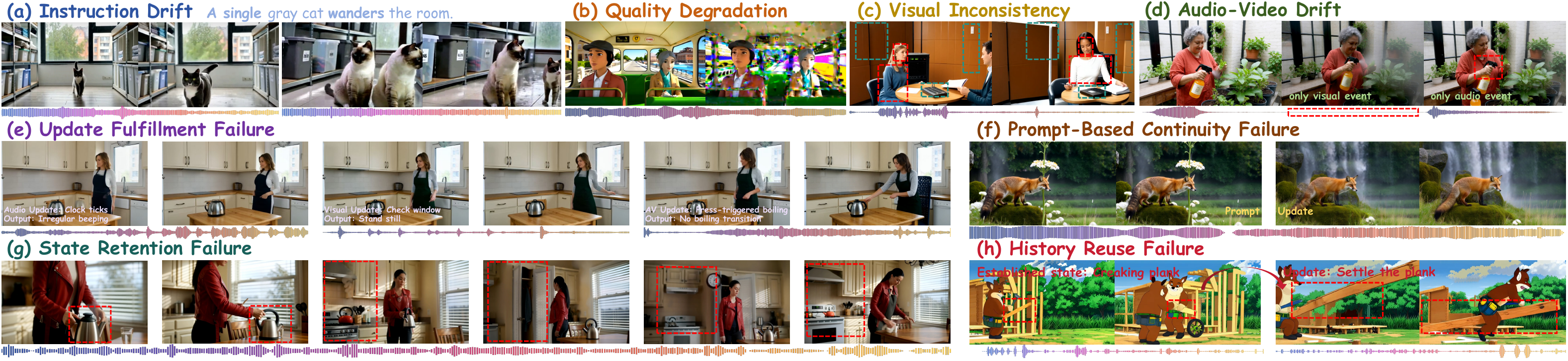}
\caption{Qualitative failure cases. Common failure modes in representative streaming generation models.}
\label{fig:qualitative_results}
\end{figure*}

\noindent\textbf{Qualitative failure analysis.}
As illustrated in \cref{fig:qualitative_results}, we identify 8 common failure modes in streaming generation. 
During progressive generation, models exhibit instruction drift, quality degradation, visual inconsistency, and audio-video drift. 
During interaction, they struggle to execute runtime updates, retain previously established states, or reuse earlier information. Additionally, abrupt temporal discontinuities frequently emerge at prompt-update boundaries. Collectively, these observations underscore the bottlenecks in long-horizon stability, interactive response, state retention and reuse, and prompt-based continuity.

\subsection{Human Alignment}
We conduct blind pairwise comparisons on a balanced subset of 60 cases covering both tracks and all 13 systems.
Visual and audio quality (\textit{VQ} and \textit{AQ}) are validated across both tracks, whereas instruction adherence metrics are evaluated on the progressive track, and interactive response and state retention and reuse metrics on the interactive track.
For each matched input and metric, three annotators compare anonymized outputs with randomized left-right order and ties allowed.
We aggregate human preferences into model-level win rates and report their Spearman correlation with the corresponding direction-adjusted benchmark scores.
As shown in \cref{tab:human_alignment}, all 17 metrics show strong model-level human alignment, with correlations ranging from 0.77 to 0.94.

\begin{table}[h]
\centering
\scriptsize
\setlength{\tabcolsep}{1.2pt}
\begin{tabular*}{\columnwidth}{@{\extracolsep{\fill}}l*{17}{c}@{}}
\toprule
Metric & VQ & AQ & VIF & AIF & VID & AID & VUF & AUF & PVC & PAC & PVUAR & PAUAR & PVRL & PARL & VSR & ASR & HDF \\
\midrule
Spearman $\rho$ & 0.88 & 0.90 & 0.91 & 0.87 & 0.83 & 0.84 & 0.89 & 0.91 & 0.94 & 0.90 & 0.90 & 0.91 & 0.77 & 0.81 & 0.86 & 0.89 & 0.80 \\
\bottomrule
\end{tabular*}
\caption{Human alignment of selected evaluation metrics. We report Spearman rank correlations between per-model human win rates and the corresponding direction-adjusted benchmark scores.}
\label{tab:human_alignment}
\end{table}

\section{Insights and Future Directions}

Our evaluation identifies stateful streaming as the central paradigm for future audio-video generation, where models should maintain a persistent world state, update it dynamically, and render each sequence with minimal latency.

\noindent\textbf{Maintaining long-horizon stability.}
Long-horizon generation should be treated as maintaining a persistent state rather than simply extending sequence duration. Future methods should preserve perceptual quality, instruction adherence, subject appearance, and audio-video alignment throughout the stream. Training and inference paradigms should therefore jointly optimize absolute performance and long-horizon stability throughout the dynamic unbounded generation process, instead of clip-level objectives.

\noindent\textbf{Advancing reliable interaction.}
Each runtime update should operate on the current world state rather than generating an independent world.
The streaming model must enable modifying specified audio-video content while preserving elements that remain valid and reusing relevant information from earlier interactions, facilitating rapid and seamless state transitions. This requires selective state-update mechanisms that jointly optimize update achievement, state retention and reuse, and response latency to ensure robust free-form interaction.

\noindent\textbf{Developing joint audio-video streaming.}
To construct a persistent world, current clip-level generation methods cannot fully resolve the challenges of real-time interaction. Future systems should generate audio and video jointly and incrementally from a shared and continuously updated latent state. This native architectural design must be coupled with system-level optimizations, such as state caching and coordinated audio-video scheduling, to minimize latency while strictly maintaining synchronization.

Together, these directions shift the field from discrete video clip generation toward continuously evolving audio-video world models that can be maintained, updated, and interacted with in real time.

\section{Conclusion}

We present StreamAV-Bench, a comprehensive benchmark for long-horizon and interactive streaming audio-video generation. StreamAV-Bench contains 320 expert-verified scenarios organized into progressive and interactive tracks to evaluate continuous generation and runtime updates. Its 32-dimensional evaluation framework measures quality and alignment, streaming continuity and efficiency, instruction adherence, long-horizon stability, interactive response, and state retention and reuse. By evaluating 13 representative native joint audio-video systems and cascaded generation pipelines under a unified evaluation protocol, StreamAV-Bench supports systematic comparison and reveals common failure modes that current benchmarks cannot capture. We believe StreamAV-Bench can serve as a standardized testbed for advancing long-horizon, interactive, and efficient streaming audio-video generation.

\bibliography{streamavbench2027}
\bibliographystyle{iclr2027_conference}

\ifdefined\streamavcombined
\let\streamavenddocument\relax
\else
\def\streamavenddocument{\end{document}}
\fi
\streamavenddocument

\clearpage
\makeatletter
\let\maketitle\savedmaketitle
\let\@maketitle\savediclr@maketitle
\let\thanks\savedthanks
\makeatother
\ifdefined\streamavcombined
\else
\pdfminorversion=7
\documentclass{article}
\usepackage{iclr2027_conference,times}
\usepackage{graphicx}
\usepackage{amsmath}
\usepackage{booktabs}
\usepackage{array}
\usepackage{multirow}
\usepackage{adjustbox}
\usepackage{float}
\usepackage{listings}
\usepackage[most]{tcolorbox}
\usepackage{hyperref}
\usepackage{xurl}
\usepackage{xspace}
\usepackage[capitalize]{cleveref}

\makeatletter
\DeclareRobustCommand\onedot{\futurelet\@let@token\@onedot}
\def\@onedot{\ifx\@let@token.\else.\null\fi\xspace}
\def\eg{\emph{e.g}\onedot} \def\Eg{\emph{E.g}\onedot}
\def\ie{\emph{i.e}\onedot} \def\Ie{\emph{I.e}\onedot}
\makeatother

\newtcblisting{promptbox}[1]{
  enhanced,
  breakable,
  listing only,
  colback=gray!4,
  colframe=black!55,
  boxrule=0.5pt,
  arc=1.5pt,
  left=4pt,
  right=4pt,
  top=3pt,
  bottom=3pt,
  title={#1},
  fonttitle=\bfseries\small,
  listing options={
    basicstyle=\ttfamily\scriptsize,
    breaklines=true,
    breakatwhitespace=true,
    columns=fullflexible,
    keepspaces=true,
    showstringspaces=false,
    literate={—}{{---}}1 {–}{{--}}1 {×}{{$\times$}}1
  }
}

\renewcommand{\topfraction}{0.9}
\renewcommand{\dbltopfraction}{0.9}
\renewcommand{\textfraction}{0.08}
\renewcommand{\floatpagefraction}{0.8}
\renewcommand{\dblfloatpagefraction}{0.85}
\iclrfinalcopy
\begin{document}
\fi

\title{StreamAV-Bench: A Comprehensive Benchmark for Streaming Audio-Video
Generation\\Supplementary Material}
\author{
Kaiqi Liu\textsuperscript{1,2} \quad
Haoxuan Zeng\textsuperscript{2} \quad
Jingqi Liu\textsuperscript{1,2} \quad
Jiacong Fang\textsuperscript{1,2} \quad
Ziqi Cai\textsuperscript{2} \\
Yunyao Mao\textsuperscript{3} \quad
Henglin Liu\textsuperscript{4} \quad
Yu Sheng\textsuperscript{5} \quad
Shuchen Weng\textsuperscript{1,2}\thanks{Corresponding authors.} \quad
Boxin Shi\textsuperscript{2*} \\
\textsuperscript{1}BAAI \quad
\textsuperscript{2}PKU \quad
\textsuperscript{3}Kling \quad
\textsuperscript{4}THU \quad
\textsuperscript{5}USTC
}

\raggedbottom
\maketitle
\fancyhead{}
\appendix
\setcounter{figure}{0}
\setcounter{table}{0}
\setcounter{equation}{0}
\renewcommand{\thefigure}{A\arabic{figure}}
\renewcommand{\thetable}{A\arabic{table}}
\renewcommand{\theequation}{A\arabic{equation}}
\renewcommand{\theHfigure}{A\arabic{figure}}
\renewcommand{\theHtable}{A\arabic{table}}
\renewcommand{\theHequation}{A\arabic{equation}}
\crefname{figure}{Figure}{Figures}
\Crefname{figure}{Figure}{Figures}

\section{Benchmark Details}
\label{sec:supp_benchmark}
\suppressfloats[t]

\subsection{Task Settings}

Both tracks contain 160 scenarios and generate 180-second audio-video outputs
from a single rollout. The progressive track evaluates instruction adherence
and long-horizon generation stability using a 180-second sequence generated
from a single prompt describing a global state. The interactive track
evaluates interactive response alongside state retention and reuse using an
ordered prompt sequence, where the initial prompt defines the starting
scenario and five subsequent prompts are provided every 30 seconds to update
visual content, audio content, or both. For both tracks, we additionally
evaluate the first 60, 120, and 180 seconds of the same rollout.

\subsection{Benchmark Construction}

\paragraph{Taxonomy design and scenario allocation.}
We construct 160 scenario themes and use every theme exactly once in each
track. The two track instances of a theme share its
semantic scope but are independently realized under their respective temporal
protocols. Before script construction, each scenario is assigned a scene domain,
audio domain, subject category, visual style, and entity, activity, and audio
complexity levels. These allocations balance semantic coverage and content
complexity. For the interactive track, the five runtime updates are further
distributed across interaction type, update modality, and temporal dependency.

\paragraph{Script construction and interval annotation.}
For each allocated scenario, annotators develop a structured script describing
the scene, visual style, primary subjects, activities, and sound sources, with
the assigned complexity levels controlling their number and organization.
Progressive scripts define coherent, open-ended audio-video processes that can
continue throughout a 180-second rollout. Interactive scripts instead jointly
plan the initial world and all five runtime updates as a single state trajectory,
rather than constructing the updates independently.

Each runtime update expresses one primary interaction intent and is annotated
with its interaction type, update modality, temporal dependency, and, when
applicable, the earlier update on which it depends. The complete sequence is
required to be causally ordered and state-consistent: established subjects,
objects, activities, sounds, and states persist until an update explicitly
changes them. An adjacent update depends on the immediately preceding runtime
update, whereas a long-range update depends on a non-adjacent earlier runtime
update.

\paragraph{Prompt generation.}
Based on the structured scripts and interval annotations,
GPT-5.4~\cite{openai_2026_gpt54} generates natural-language prompts. Progressive
scenarios yield one self-contained global prompt. Interactive scenarios yield a
self-contained initial prompt followed by five concise runtime updates for
subsequent intervals. All prompts follow a common, generation-oriented style.

\paragraph{Checklist generation and expert verification.}
Following prompt generation, GPT-5.4 uses fixed checklist-construction
templates to derive modality-specific visual and audio checklists. Each
interactive runtime update defines its evaluation target; joint audio-video
updates are split into visual and audio targets while preserving their original
semantics. For every adjacent or long-range update, the corresponding earlier
prompt is associated with the target.

We assign two domain experts as independent reviewers and a third as an
adjudicator. The reviewers examine the scripts, prompts, checklists, update
targets, and temporal dependencies. They verify consistency with the assigned
taxonomy and complexity design, update order, update modality, and temporal
dependencies, and assess generation feasibility, audio-video plausibility,
content coverage, temporal and cross-update state consistency, and evaluation
consistency. Cases requiring correction are revised and reviewed again. The
adjudicator resolves any
disagreements, producing the final expert-verified set of 320 scenarios. All
evaluation materials are finalized before model evaluation.

\subsection{Benchmark Statistics}

\Cref{tab:supp_taxonomy} reports the semantic distribution over the 320
scenarios. StreamAV-Bench spans 8 scene and 5 audio domains, 5 subject
categories, and 4 visual styles, including 192 photorealistic and 128
non-photorealistic scenarios. Subject categories are assigned by the primary
subject of each scenario, while audio domains identify the principal audio
requirement.

\begin{table}[H]
\centering
\small
\setlength{\tabcolsep}{4pt}
\begin{tabular}{p{0.16\textwidth}p{0.76\textwidth}}
\toprule
Dimension & Distribution \\
\midrule
Scene domain &
Daily Life 48; Work \& Creation 44; Social 44; Performance 40; Sports 40;
Transportation 36; Nature 36; Fictional 32 \\
Audio domain &
Speech \& Vocalization 80; Foley \& Object Sound 64; Mechanical Sound 48;
Environmental Ambience 64; Music 64 \\
Subject category &
Human 173; Animal/Creature 48; Object 35; Vehicle 24; Robot 40 \\
Visual style &
Photorealistic 192; Anime/Cartoon 48; 3D/CG 48; Painting/Illustration 32 \\
\bottomrule
\end{tabular}
\caption{Semantic distribution over the 320 scenarios.}
\label{tab:supp_taxonomy}
\end{table}

\paragraph{Content complexity.}
Both tracks characterize content complexity along three dimensions.
\textit{Entity complexity} measures the number of primary subjects that must
remain individually recognizable: Single contains one primary subject, Multi
contains two or three, and Dense contains four or more. Primary subjects may be
people, animals, vehicles, robots, or other persistent focal entities rather
than all visible objects. Each track contains 56 Single, 72 Multi, and 32 Dense
scenarios. \textit{Activity complexity} characterizes activity structure:
Atomic contains one independent primary activity, Relational contains one
dominant activity involving an explicit subject--subject or subject--object
relation, and Concurrent contains multiple distinguishable activity threads
occurring simultaneously. Each track contains 40 Atomic, 80 Relational, and 40
Concurrent scenarios. \textit{Audio complexity} characterizes the number and
temporal organization of salient sound sources: Single-Source contains one
primary active source, Multi-Source contains multiple sources with little or no
substantial overlap, and Overlapping contains multiple distinguishable sources
active simultaneously. Each track contains 48 Single-Source, 64 Multi-Source,
and 48 Overlapping scenarios. For interactive scenarios, these labels describe
the initial world established before runtime updates.

\paragraph{Update complexity.}
The interactive track further characterizes each runtime update by what changes,
which modality changes, and how much interaction history is required.
\textit{Interaction type} distinguishes agent actions, entity-state changes,
environment changes, and audio changes; \textit{update modality} distinguishes
video-only, audio-only, and joint audio-video updates; and \textit{temporal
dependency} distinguishes independent updates from those relying on the
immediately preceding or a non-adjacent earlier update. Across 800 updates,
interaction types comprise 440 agent actions, 160 entity-state changes, 80
environment changes, and 120 audio changes; update modalities comprise 360
video-only, 160 audio-only, and 280 joint audio-video updates; temporal
dependencies comprise 440 independent, 240 adjacent, and 120 long-range
updates.

\section{Evaluation Framework Details}
\label{sec:supp_metrics}

Our evaluation framework combines tailored expert models with Multimodal Large
Language Model (MLLM) assessments guided by case-specific checklists. The 32
metrics are organized into six categories. The progressive track primarily
measures instruction adherence and long-horizon
stability, while the interactive track focuses on interactive response and
state retention and reuse. Shared metrics evaluate streaming continuity and
efficiency as well as quality and alignment on both tracks. For each metric, we
summarize the evaluated property, evaluator, score construction, and direction.
Unless stated otherwise, observations are first aggregated within each case and
then averaged uniformly across cases.

\subsection{Quality and Alignment}

We assess generated audio-video content along three complementary dimensions:
visual quality, audio quality, and cross-modal alignment. The six metrics in
this category are computed over six non-overlapping 30-second intervals and
then averaged within each case.

\paragraph{(1) Visual Aesthetics (VA).}
VA measures perceptual visual appeal using the LAION Aesthetic
Predictor~\cite{laion_2022_aesthetic}. Interval scores are normalized to
$[0,1]$ and averaged; higher is better.

\paragraph{(2) Visual Quality (VQ).}
VQ evaluates visual fidelity, subject integrity, motion naturalness, and
artifact control using Gemini 3.1 Pro~\cite{google_2026_gemini31pro}. The four
dimensions are averaged on the original 1--5 scale; higher is better.

\paragraph{(3) Production Quality (PQ).}
PQ measures perceptual and production fidelity using the production-quality
head of Audiobox Aesthetics~\cite{tjandra_2025_audiobox}. We report the original
evaluator score; higher is better.

\paragraph{(4) Audio Quality (AQ).}
AQ evaluates audio naturalness and artifact control using Gemini 3.1 Pro. The
two dimensions are averaged on a 1--5 scale; higher is better.

\paragraph{(5) Audio-Video Alignment (AVAlign).}
AVAlign measures semantic correspondence between generated audio and visual
content using cosine similarity in the shared ImageBind embedding
space~\cite{girdhar_2023_imagebind}. Similarities are averaged over intervals;
higher is better.

\paragraph{(6) Audio-Video Synchronization (AVSync).}
AVSync measures temporal synchronization using the absolute audio-video offset
predicted by Synchformer~\cite{iashin_2024_synchformer}. Offsets are averaged
over intervals and reported in seconds; lower is better.

\subsection{Streaming Continuity and Efficiency}

\paragraph{(7) Native Boundary Continuity (NBC).}
NBC measures technical continuity at observable model-native generation-unit
boundaries by detecting black frames, repeated frames, freezing, and
flashes. Building on the transition-artifact analysis of
LongAV-Compass~\cite{liu_2026_longavcompass}, NBC expands this evaluation to
model-native generation-unit boundaries. Let $r_b$, $r_d$, $t_f$, and $n_f$
denote the black-frame ratio, repeated-frame ratio, maximum freeze duration in
seconds, and flash count, respectively. The boundary score is
\begin{equation}
\mathrm{NBC}=
\operatorname{clip}\!\left(
100-\left(70r_b+45r_d+12t_f+8n_f\right),0,100
\right).
\end{equation}
Scores are averaged over boundaries and cases from both tracks; higher is
better.

\paragraph{(8) Frames per Second (FPS).}
FPS measures generated video frames per unit generation time under each
system's default recommended configuration; higher is better.

\paragraph{(9) Time-to-First-Chunk (TTFC).}
TTFC measures the elapsed time from receiving the generation request to
producing the first output chunk; lower is better.

\subsection{Instruction Adherence}

For each progressive case, fixed visual and audio checklists derived from the
global prompt define the requirements used by all instruction-adherence
metrics.

\paragraph{(10--11) Video and Audio Instruction Fulfillment (VIF/AIF).}
VIF and AIF measure how completely the visual and audio requirements are
realized over the full output. VIF covers scene, visual style, subjects, and
activity, while AIF covers sound source, sound content, and temporal relations.
Gemini 3.1 Pro scores each applicable criterion on a 1--5 completion scale, and
criterion scores are averaged within each case; higher is better.

\paragraph{(12--13) Video and Audio Instruction Drift (VID/AID).}
VID and AID measure changes in visual and audio instruction adherence between
the first and last intervals using the same fixed criteria. Following the
endpoint comparison of HeliosBench~\cite{yuan_2026_helios}, for each matched
criterion $c$ we compute
\begin{equation}
D_c=\left|s_{c}^{(6)}-s_{c}^{(1)}\right|.
\end{equation}
VID and AID average $D_c$ over matched visual and audio criteria, respectively.
Both lie in $[0,4]$; lower is better.

\subsection{Long-Horizon Stability}

The progressive track evaluates quality drift, visual consistency, and
audio-video drift over the full generation horizon. Endpoint-based drift metrics
compare the first and last 30-second intervals.

\paragraph{(14--17) Quality Drift.}
VA-D, VQ-D, PQ-D, and AQ-D measure the absolute endpoint change of their
corresponding shared quality metrics:
\begin{equation}
Q\text{-D}=\left|q^{(6)}-q^{(1)}\right|.
\end{equation}
Differences are computed per case and then averaged; lower is better.

\paragraph{(18--19) Subject Consistency (SC) and Background Consistency (BC).}
SC and BC measure the temporal consistency of subject appearance and background
scenes, respectively, following the official VBench-Long
protocol~\cite{huang_2025_vbenchpp}. Scores are reported on the original
evaluator scale; higher is better.

\paragraph{(20--21) Audio-Video Drift (AVAlign-D/AVSync-D).}
AVAlign-D and AVSync-D measure absolute endpoint changes in audio-video semantic
correspondence and temporal synchronization. They use the original AVAlign and
AVSync units; lower is better.

\subsection{Interactive Response}

The interactive track evaluates whether systems realize runtime updates
correctly, smoothly, and promptly. Metrics are computed at the five update
boundaries at 30, 60, 90, 120, and 150 seconds. VUF, AUF, PVUAR, PAUAR, PVRL,
and PARL follow update-modality applicability, while PVC and PAC evaluate all
five boundaries.

\paragraph{(22--23) Video and Audio Update Fulfillment (VUF/AUF).}
VUF and AUF measure how completely each update's visual and audio targets are
realized. Gemini 3.1 Pro assigns a 1--5 fulfillment score to each applicable
target, and update scores are averaged within each case; higher is better.

\paragraph{(24) Prompt-Based Visual Continuity (PVC).}
PVC measures visual continuity across prompt-update boundaries. It combines a
deterministic boundary score for black frames, repetition, freezing, and flashes
with a Gemini 3.1 Pro judgment of unintended breaks, deformation, and
disappearance, while allowing requested changes. The deterministic component
applies the NBC boundary-artifact scoring function at each prompt-update
boundary. Let $S_{\mathrm{B}}$ denote the resulting 0--100 boundary score. We
map it to the 1--5 scale as $S_{\mathrm{alg}}=1+4S_{\mathrm{B}}/100$ and fuse
it with the MLLM score:
\begin{equation}
S_{\mathrm{PVC}}=0.70S_{\mathrm{alg}}+0.30S_{\mathrm{MLLM}}.
\end{equation}
Scores are averaged over update boundaries; higher is better.

\paragraph{(25) Prompt-Based Audio Continuity (PAC).}
PAC measures audio continuity across prompt-update boundaries using Gemini 3.1
Pro. It penalizes unintended dropouts, hard cuts, repetition, clicks, clipping,
and abrupt volume changes while allowing requested changes. Scores use a 1--5
scale and are averaged over boundaries; higher is better.

\paragraph{(26--27) Prompt-Based Visual and Audio Update Achievement Rates (PVUAR/PAUAR).}
PVUAR and PAUAR measure the fractions of applicable visual and audio updates
whose core targets are clearly realized. A target counts as achieved only when
the evaluator identifies a clear realization time; merely beginning a change is
insufficient. The case-level visual rate is
\begin{equation}
\mathrm{PVUAR}=
\frac{\#\{\text{applicable visual updates with an achieved target}\}}
{\#\{\text{applicable visual updates}\}},
\end{equation}
and PAUAR is defined analogously over applicable audio updates. Rates lie in
$[0,1]$; higher is better.

\paragraph{(28--29) Prompt-Based Visual and Audio Response Latency (PVRL/PARL).}
Conditional on target achievement, PVRL and PARL measure the media time from the
start of the post-update interval to the first clear realization of the visual
and audio targets, respectively. Latencies are averaged over successful updates
within each case and then over cases with at least one achieved update. They are
reported in seconds on the generated content timeline; lower is better.

\subsection{State Retention and Reuse}

This category evaluates whether systems preserve states that should remain
unchanged and correctly reuse relevant information from earlier interactions.
HDF is computed only when its required source state is established.

\paragraph{(30) Visual State Retention (VSR).}
VSR measures whether subject appearance and the surrounding environment remain
consistent across an update after accounting for requested visual changes.
Gemini 3.1 Pro assigns 1--5 retention scores; higher is better.

\paragraph{(31) Audio State Retention (ASR).}
ASR measures whether recurring music, ambience, and other persistent sounds
remain consistent when they are not targeted by the update. Gemini 3.1 Pro
assigns a 1--5 retention score; higher is better.

\paragraph{(32) History-Dependency Following (HDF).}
HDF measures whether an adjacent or long-range update is correctly executed
using the state established by its annotated earlier interaction. Gemini 3.1
Pro assigns a 1--5 score over established source states. Adjacent and long-range
instances are combined within each case in proportion to their applicable
updates; higher is better. Source-state establishment rates are reported in
\cref{tab:supp_conditional_coverage}.

\section{Implementation Details}

\subsection{Evaluated Systems}

We evaluate 13 systems across two generation paradigms, as summarized in
\Cref{tab:supp_systems}.
PixVerse R1~\cite{pixverse_2026_r1}, HappyOyster~\cite{alibaba_2026_happyoyster},
and OmniForcing~\cite{su_2026_omniforcing} are native joint streaming
audio-video models.
Cascaded pipelines pair HunyuanVideo-Foley~\cite{shan_2025_hunyuanfoley} with
native streaming video models:
Odyssey-2~\cite{odyssey_2025_odyssey2},
Self-Forcing~\cite{huang_2025_selfforcing},
LongLive~\cite{yang_2025_longlive},
Rolling-Forcing~\cite{liu_2025_rollingforcing},
Deep Forcing~\cite{yi_2025_deepforcing},
MemFlow~\cite{ji_2025_memflow},
Causal-Forcing~\cite{zhu_2026_causalforcing},
Helios~\cite{yuan_2026_helios},
SWIFT~\cite{tan_2026_swift}, and
IAMFlow~\cite{liu_2026_iamflow}.
Open-source systems use their official checkpoints and inference pipelines on
NVIDIA H800 GPUs. Commercial systems are accessed via their official web
interfaces. All models are deployed with their default recommended generation
configurations.
Because the official web interfaces do not expose model-native generation-unit
boundaries or reliable generation-time traces, NBC, FPS, and TTFC are
unavailable for PixVerse R1, HappyOyster, and Odyssey-2.

\begin{table}[H]
\centering
\small
\begin{tabular}{lll}
\toprule
System & Generation paradigm & Access \\
\midrule
PixVerse R1 & Native joint & Web \\
HappyOyster & Native joint & Web \\
OmniForcing & Native joint & Open \\
\midrule
Odyssey-2 & Cascaded & Web \\
Self-Forcing & Cascaded & Open \\
LongLive & Cascaded & Open \\
Rolling-Forcing & Cascaded & Open \\
Deep Forcing & Cascaded & Open \\
MemFlow & Cascaded & Open \\
Causal-Forcing & Cascaded & Open \\
Helios & Cascaded & Open \\
SWIFT & Cascaded & Open \\
IAMFlow & Cascaded & Open \\
\bottomrule
\end{tabular}
\caption{Evaluated systems. ``Native joint'' denotes native joint streaming
audio-video generation, and ``Cascaded'' denotes a cascaded streaming
audio-video pipeline.}
\label{tab:supp_systems}
\end{table}

\subsection{Interactive Continuation Protocol}

For Self-Forcing~\cite{huang_2025_selfforcing},
Rolling-Forcing~\cite{liu_2025_rollingforcing}, Deep
Forcing~\cite{yi_2025_deepforcing},
Causal-Forcing~\cite{zhu_2026_causalforcing}, and
OmniForcing~\cite{su_2026_omniforcing}, we use a shared runtime-update protocol
based on the continuation strategies of
Infinity-RoPE~\cite{yesiltepe_2026_infinityrope} and
LongLive~\cite{yang_2025_longlive}. The initial prompt starts the rollout;
subsequent prompts become active at fixed 30-second boundaries while previously
generated frames remain unchanged. At each boundary, conditioning switches to
the active prompt and generation continues from the existing history.

\subsection{Cascaded Audio Generation}

Each cascaded system pairs its streaming video generator with
HunyuanVideo-Foley~\cite{shan_2025_hunyuanfoley}. Audio generation follows the
same prompt schedule as the video stream, with each generated audio segment
conditioned on the prompt active at the corresponding time.

\subsection{Generation Controls}

To ensure a consistent evaluation protocol, all systems process identical
benchmark inputs: a single global prompt for the progressive track, and an
ordered prompt sequence with fixed 30-second intervals for the interactive
track. Outputs are converted to a common audio-video format before evaluation.

\section{Additional Experimental Results}

\subsection{Initial-World Alignment}

The core interactive evaluation focuses on runtime-update response, state
retention, and reuse. To complement these core dimensions, we additionally
evaluate whether each system establishes the audio-video world requested by the
initial prompt before the first runtime update. We apply the same VIF/AIF
framework to checklists derived from the initial prompt over the first interval.
Scores are averaged over applicable criteria within each case and then uniformly
across the 160 interactive cases. \Cref{tab:supp_initial_alignment} reports the
resulting per-system scores.

\begin{table}[H]
\centering
\small
\begin{tabular}{lcc}
\toprule
System & Initial VIF$\uparrow$ & Initial AIF$\uparrow$ \\
\midrule
\multicolumn{3}{l}{\emph{Native joint streaming audio-video models}} \\
PixVerse R1 & 2.954 & 2.467 \\
HappyOyster & \textbf{3.364} & 2.492 \\
OmniForcing & 2.527 & 2.299 \\
\midrule
\multicolumn{3}{l}{\emph{Cascaded streaming audio-video pipelines}} \\
Odyssey-2 & 3.084 & 2.643 \\
Self-Forcing & 3.011 & 2.638 \\
LongLive & 3.130 & 2.675 \\
Rolling-Forcing & 3.077 & 2.686 \\
Deep Forcing & 3.083 & 2.654 \\
MemFlow & 3.117 & 2.645 \\
Causal-Forcing & 2.930 & 2.590 \\
Helios & 2.908 & 2.710 \\
SWIFT & \underline{3.131} & \textbf{2.753} \\
IAMFlow & 3.100 & \underline{2.722} \\
\bottomrule
\end{tabular}
\caption{Initial-world alignment results on the interactive track. Initial VIF
and Initial AIF reuse the VIF/AIF 1--5 scale. The best and second-best results
are highlighted in \textbf{bold} and \underline{underlined}.}
\label{tab:supp_initial_alignment}
\end{table}

HappyOyster attains the highest Initial VIF, whereas SWIFT attains the highest
Initial AIF. No single system leads both modalities, indicating complementary
strengths in establishing the requested starting world.

\subsection{Update Achievement and Response Latency}

Update achievement and response latency capture complementary aspects of
interactive performance: the former measures how often a requested change is
realized, whereas the latter measures when a successful realization first
appears on the generated content timeline. Because PVRL and PARL are conditioned
on achieved targets, lower latency does not by itself imply stronger overall
responsiveness.
\Cref{tab:supp_tar_tal} contrasts the system with the highest update
achievement rates against those attaining the lowest visual and audio response
latencies. HappyOyster obtains \textit{PVUAR} and \textit{PAUAR} scores of
0.518 and 0.488, with corresponding \textit{PVRL} and \textit{PARL} values of
14.484 and 13.621 seconds. SWIFT achieves a lower \textit{PVRL} of 10.340
seconds but a \textit{PVUAR} of 0.246, while OmniForcing achieves the lowest
\textit{PARL} of 7.925 seconds but a \textit{PAUAR} of 0.316.

\begin{table}[H]
\centering
\small
\begin{tabular}{lcccc}
\toprule
System & PVUAR$\uparrow$ & PVRL$\downarrow$ & PAUAR$\uparrow$ & PARL$\downarrow$ \\
\midrule
HappyOyster & 0.518 & 14.484 & 0.488 & 13.621 \\
SWIFT & 0.246 & 10.340 & 0.276 & 12.094 \\
OmniForcing & 0.053 & 13.295 & 0.316 & 7.925 \\
\bottomrule
\end{tabular}
\caption{Prompt-based visual and audio update achievement rates and response
latencies for three representative systems. PVRL and PARL are media-time
latencies measured on generated content over achieved targets.}
\label{tab:supp_tar_tal}
\end{table}

HappyOyster realizes a larger fraction of requested updates, whereas the lower
visual or audio latency of SWIFT and OmniForcing is measured over a substantially
smaller successful subset. The failure-penalized comparison in
\cref{sec:supp_adjusted_diagnostics} evaluates achievement and latency jointly.

\subsection{History-Dependency Source Establishment}
\label{sec:supp_conditional_coverage}

History-dependency following can be evaluated only when the required state is
present in its annotated earlier interaction.
\Cref{tab:supp_conditional_coverage} reports source-state establishment rates
for adjacent (HDF-A) and long-range (HDF-L) dependencies, together with the
corresponding HDF scores over established sources.

\begin{table}[H]
\centering
\small
\setlength{\tabcolsep}{5pt}
\begin{tabular}{lcccc}
\toprule
System & \multicolumn{2}{c}{Source Establishment} &
\multicolumn{2}{c}{Conditional HDF} \\
\cmidrule(lr){2-3}\cmidrule(lr){4-5}
& Adj.$\uparrow$ & Long$\uparrow$ & Adj.$\uparrow$ & Long$\uparrow$ \\
\midrule
PixVerse R1     & 0.610 & 0.653 & 2.508 & 2.435 \\
HappyOyster     & \textbf{0.741} & \underline{0.715} & \textbf{2.885} & \textbf{2.718} \\
OmniForcing     & 0.323 & 0.358 & \underline{2.641} & 2.231 \\
Odyssey-2       & 0.459 & 0.453 & 2.631 & 2.489 \\
Self-Forcing    & 0.568 & 0.606 & 2.445 & 2.126 \\
LongLive        & 0.612 & 0.655 & 2.212 & 2.167 \\
Rolling-Forcing & 0.607 & 0.595 & 2.614 & 2.138 \\
Deep Forcing    & 0.560 & 0.555 & 2.278 & 2.088 \\
MemFlow         & 0.571 & 0.580 & 2.377 & \underline{2.500} \\
Causal-Forcing  & 0.537 & 0.559 & 2.363 & 2.357 \\
Helios          & 0.536 & 0.544 & 2.491 & 2.357 \\
SWIFT           & \underline{0.616} & \textbf{0.716} & 2.265 & 2.348 \\
IAMFlow         & 0.565 & 0.610 & 2.496 & 2.216 \\
\bottomrule
\end{tabular}
\caption{History-dependency source-state establishment rates and conditional
HDF scores. Establishment rates are reported on $[0,1]$, while HDF-A and HDF-L
scores use the original 1--5 scale over established sources. The best and
second-best results are highlighted in \textbf{bold} and
\underline{underlined}.}
\label{tab:supp_conditional_coverage}
\end{table}

HappyOyster leads adjacent source establishment and both conditional HDF
scores, while SWIFT attains the highest long-range source establishment.
OmniForcing ranks second in conditional HDF-A despite the lowest adjacent
establishment rate, showing that source establishment and dependency following
capture distinct failure modes.

\subsection{Success- and Coverage-Adjusted Diagnostics}
\label{sec:supp_adjusted_diagnostics}

We define success-adjusted response latencies (SA-PVRL/SA-PARL) and
coverage-adjusted HDF scores (CA-HDF-A/CA-HDF-L) by assigning fixed penalties
to unachieved updates and unestablished dependency sources.

For modality $m\in\{\mathrm{V},\mathrm{A}\}$, let
$\mathcal{U}_{c}^{m}$ be the applicable updates in case $c$,
$a_{c,u}^{m}$ indicate whether the target is achieved, and
$\ell_{c,u}^{m}$ be its response latency when achieved. Let
$\mathcal{C}_{m}$ denote cases containing at least one such update. With the
fixed observation horizon $T=30$ seconds, we define
\begin{equation}
\widetilde{\ell}_{c,u}^{m} =
\begin{cases}
\ell_{c,u}^{m}, & a_{c,u}^{m}=1,\\
T, & a_{c,u}^{m}=0,
\end{cases}
\qquad
L_{\mathrm{SA}}^{m}
=
\frac{1}{|\mathcal{C}_{m}|}
\sum_{c\in\mathcal{C}_{m}}
\frac{1}{|\mathcal{U}_{c}^{m}|}
\sum_{u\in\mathcal{U}_{c}^{m}}
\widetilde{\ell}_{c,u}^{m}.
\end{equation}
This yields success-adjusted PVRL and PARL (SA-PVRL/SA-PARL), for which
unachieved targets receive the observation-window cap rather than being
excluded. These remain media-time latencies on the generated content timeline.

For a dependency class
$d\in\{\mathrm{HDF\text{-}A},\mathrm{HDF\text{-}L}\}$, let
$z_{c,j}^{d}$ indicate whether the source state for check $j$ is established
and let $s_{c,j}^{d}\in[1,5]$ be the corresponding conditional score. We use
$\mathcal{J}_{c}^{d}$ for the checks in case $c$ and $\mathcal{C}_{d}$ for
cases containing at least one such check, and define
\begin{equation}
\widetilde{s}_{c,j}^{d} =
\begin{cases}
(s_{c,j}^{d}-1)/4, & z_{c,j}^{d}=1,\\
0, & z_{c,j}^{d}=0,
\end{cases}
\qquad
S_{\mathrm{CA}}^{d}
=
\frac{1}{|\mathcal{C}_{d}|}
\sum_{c\in\mathcal{C}_{d}}
\frac{1}{|\mathcal{J}_{c}^{d}|}
\sum_{j\in\mathcal{J}_{c}^{d}}
\widetilde{s}_{c,j}^{d}.
\end{equation}
This yields CA-HDF-A and CA-HDF-L for the two dependency classes. Both scores
lie on $[0,1]$.

\begin{table}[H]
\centering
\small
\setlength{\tabcolsep}{4.5pt}
\begin{tabular}{lcccc}
\toprule
System & SA-PVRL$\downarrow$ & SA-PARL$\downarrow$ &
CA-HDF-A$\uparrow$ & CA-HDF-L$\uparrow$ \\
\midrule
PixVerse R1     & \underline{24.011} & \textbf{20.855} & 0.232 & \underline{0.234} \\
HappyOyster     & \textbf{21.991} & \underline{21.861} & \textbf{0.354} & \textbf{0.306} \\
OmniForcing     & 29.128 & 23.046 & 0.136 & 0.109 \\
Odyssey-2       & 26.838 & 25.611 & 0.196 & 0.168 \\
Self-Forcing    & 27.081 & 24.156 & 0.206 & 0.169 \\
LongLive        & 25.575 & 25.490 & 0.186 & 0.187 \\
Rolling-Forcing & 26.355 & 24.941 & \underline{0.250} & 0.155 \\
Deep Forcing    & 26.420 & 25.024 & 0.181 & 0.150 \\
MemFlow         & 26.503 & 24.433 & 0.200 & 0.216 \\
Causal-Forcing  & 26.497 & 25.117 & 0.184 & 0.188 \\
Helios          & 26.841 & 23.712 & 0.202 & 0.181 \\
SWIFT           & 25.150 & 25.195 & 0.192 & 0.232 \\
IAMFlow         & 25.976 & 24.936 & 0.213 & 0.183 \\
\bottomrule
\end{tabular}
\caption{Sensitivity diagnostics on the interactive track.
SA-PVRL and SA-PARL are measured in seconds and assign the 30-second
observation-window cap to unachieved targets. CA-HDF metrics combine
source-state establishment with normalized conditional HDF quality. The best
and second-best results are highlighted in \textbf{bold} and
\underline{underlined}.}
\label{tab:supp_adjusted_diagnostics}
\end{table}

HappyOyster obtains the lowest SA-PVRL and the highest CA-HDF-A and CA-HDF-L,
while PixVerse R1 obtains the lowest SA-PARL.

\subsection{Temporal Evolution}

We evaluate progressive-track quality and audio-video alignment over six
consecutive 30-second intervals. For each case $c$, let $q_{c,i}$ denote its
score in interval $i$ and $t_i\in\{30,60,\ldots,180\}$ the corresponding time
in seconds. Degradation slope captures the overall direction across all
segments, while trajectory variation measures adjacent-segment fluctuations.
We first fit each case trajectory using
\begin{equation}
    \hat{\beta}_c
    =
    \frac{\sum_{i=1}^{6}(t_i-\bar{t})(q_{c,i}-\bar{q}_c)}
         {\sum_{i=1}^{6}(t_i-\bar{t})^2}.
\end{equation}
The case-level degradation slope is $-100\hat{\beta}_c$ for higher-is-better
dimensions and $100\hat{\beta}_c$ for AVSync, such that a positive value always
indicates deterioration and the unit is score change per 100 seconds. We
additionally compute case-level trajectory variation as
\begin{equation}
    \mathrm{TV}_c
    =
    \frac{1}{5}\sum_{i=1}^{5}\left|q_{c,i+1}-q_{c,i}\right|.
\end{equation}
Both statistics are then averaged uniformly across cases for each system. TV
remains in each dimension's native score unit.

Visual aesthetics (\textit{VA}), audio-video alignment (\textit{AVAlign}), and
visual quality (\textit{VQ}) degrade in 13, 12, and 10 of the 13 systems,
respectively. By contrast, production quality (\textit{PQ}), audio quality
(\textit{AQ}), and audio-video synchronization (\textit{AVSync}) exhibit mixed
directions, degrading in 9, 6, and 7 systems. Trajectory variation further
reveals instability concealed by the overall trend: Odyssey-2 and PixVerse R1
show improving \textit{VQ} and \textit{AQ} trends ($-0.056$ and $-0.019$), yet
the largest variations in their respective dimensions (0.423 and 0.508).
Meanwhile, SWIFT exhibits both the steepest \textit{AVSync} degradation (0.062)
and the largest \textit{AVSync} variation (0.536). Streaming evolution is
therefore dimension-specific: visual quality and semantic alignment degrade
most consistently, whereas audio quality and temporal synchronization often
fluctuate rather than decline monotonically.

\paragraph{Generation-length analysis.}
We compare performance at 60, 120, and 180 seconds across both tracks, using
the same 180-second rollouts. For progressive cases, the endpoint drift of
metric $m$ at horizon $h$ is
\begin{equation}
    D_{c}^{m}(h)
    =
    \left|q_{c,k(h)}^{m}-q_{c,1}^{m}\right|,
    \qquad
    k(60)=2,\;k(120)=4,\;k(180)=6.
\end{equation}
For interactive cases, the three horizons aggregate the first one, three, and
five runtime updates, respectively. Scores are first aggregated within each
case and then averaged across cases and systems. The resulting values are
reported in~\cref{tab:supp_duration_scaling}.

\begin{table}[H]
\centering
\small
\begin{tabular*}{\linewidth}{@{\extracolsep{\fill}}lccc@{\hskip 1.6em}lccc@{}}
\toprule
\multicolumn{4}{c}{Progressive track} &
\multicolumn{4}{c}{Interactive track} \\
\cmidrule(lr){1-4} \cmidrule(lr){5-8}
Metric & 60 s & 120 s & 180 s &
Metric & 60 s & 120 s & 180 s \\
\midrule
VA-D$\downarrow$      & 0.025 & 0.036 & 0.042 &
VUF$\uparrow$         & 2.347 & 2.327 & 2.347 \\
VQ-D$\downarrow$      & 0.313 & 0.347 & 0.361 &
AUF$\uparrow$         & 2.481 & 2.340 & 2.332 \\
PQ-D$\downarrow$      & 0.295 & 0.356 & 0.382 &
PVUAR$\uparrow$       & 0.233 & 0.231 & 0.243 \\
AQ-D$\downarrow$      & 0.449 & 0.460 & 0.495 &
PAUAR$\uparrow$       & 0.361 & 0.333 & 0.331 \\
AVAlign-D$\downarrow$ & 0.059 & 0.067 & 0.065 &
PVRL$\downarrow$      & 12.283 & 13.268 & 13.534 \\
AVSync-D$\downarrow$  & 0.495 & 0.521 & 0.509 &
PARL$\downarrow$      & 11.153 & 12.116 & 12.437 \\
& & & &
VSR$\uparrow$         & 2.606 & 2.619 & 2.637 \\
& & & &
ASR$\uparrow$         & 3.022 & 2.952 & 2.895 \\
\bottomrule
\end{tabular*}
\caption{Generation-length analysis at 60, 120, and 180 seconds, averaged
uniformly across the 13 systems. Progressive values report prefix endpoint
drift, while interactive values aggregate all runtime updates completed by
each horizon. PVRL and PARL are media-time latencies measured in seconds on the
generated content timeline.}
\label{tab:supp_duration_scaling}
\end{table}

Averaged across the 13 systems, the progressive track exhibits increasing
quality drift: visual aesthetics drift (\textit{VA-D}) rises from 0.025 to
0.036 and 0.042, while visual quality drift (\textit{VQ-D}) increases from
0.313 to 0.347 and 0.361. Production and audio quality drift (\textit{PQ-D} and
\textit{AQ-D}) follow the same pattern, whereas audio-video alignment and
synchronization drift (\textit{AVAlign-D} and \textit{AVSync-D}) change little
beyond 120 seconds. In the interactive track, visual update fulfillment
(\textit{VUF}) and achievement (\textit{PVUAR}) remain stable, whereas audio
update fulfillment (\textit{AUF}) decreases from 2.481 to 2.340 and 2.332, and
audio update achievement (\textit{PAUAR}) decreases from 0.361 to 0.333 and
0.331. Visual and audio response latencies (\textit{PVRL} and \textit{PARL})
both increase by approximately 1.3 seconds. Audio state retention
(\textit{ASR}) also declines from 3.022 to 2.952 and 2.895, while visual state
retention (\textit{VSR}) remains stable. Longer rollouts therefore accumulate
quality drift in progressive generation while primarily exposing audio-side
degradation and slower responses under repeated interaction.

\subsection{Content and Update Complexity Results}

We analyze the progressive track across entity, activity, and audio complexity.
Entity density produces the clearest visual degradation: visual instruction
fulfillment (\textit{VIF}) decreases from 3.090 for single-entity cases to
2.816 for dense cases, while visual quality (\textit{VQ}) decreases
monotonically from 2.760 to 2.651. Activity complexity reveals a similar visual
bottleneck, with \textit{VQ} decreasing by 0.129 from atomic to concurrent
activities without a corresponding decline in audio instruction fulfillment
(\textit{AIF}). Multi-source cases attain the lowest \textit{AIF} (2.523). In
contrast, subject consistency (\textit{SC}) and background consistency
(\textit{BC}) remain nearly unchanged across all three factors. Increasing
content complexity therefore primarily disrupts the realization of required
visual content, while subject appearance and background scenes remain
comparatively stable.

For the interactive track, visual and audio update fulfillment (\textit{VUF}
and \textit{AUF}) are comparable (2.347 vs.\ 2.332). However, visual updates
have lower achievement rates than audio updates (\textit{PVUAR}/\textit{PAUAR}:
0.243/0.331) and respond more slowly (\textit{PVRL}/\textit{PARL}:
13.534/12.437 seconds). Direct agent actions also attain higher \textit{VUF}
(2.418) than entity-state and environment changes (2.274 and 2.283). For
history-dependent updates, \textit{HDF} reaches only 2.477 for adjacent and
2.321 for long-range dependencies. The primary interaction bottlenecks
therefore lie in reliably realizing visual changes, revising established
states, and reusing earlier information.

\section{Human Alignment}

We conduct blind pairwise comparisons on a balanced subset of 60 cases covering
both tracks and all 13 systems. Visual and audio quality (\textit{VQ} and
\textit{AQ}) are validated across both tracks, whereas instruction adherence
metrics are evaluated on the progressive track, and interactive response and
state retention and reuse metrics on the interactive track. For shared metrics,
the 60 cases comprise 30 from each track; track-specific metrics use the
corresponding 30-case subset. For each metric, we sample 120 paired output
comparisons from all 13 systems using balanced random pairing. Each pair is
independently evaluated by three annotators with randomized left-right order
and ties allowed. This
produces 2,040 pairwise tasks and 6,120 individual judgments across the 17
metrics. Majority vote determines the final label. Human preferences are
aggregated into model-level win rates, with a tie contributing 0.5 to each
system. We report the Spearman correlation between human win rates and the
corresponding direction-adjusted benchmark scores.

\begin{table}[H]
\centering
\scriptsize
\setlength{\tabcolsep}{1.2pt}
\begin{tabular*}{\linewidth}{@{\extracolsep{\fill}}l*{17}{c}@{}}
\toprule
Metric & VQ & AQ & VIF & AIF & VID & AID & VUF & AUF & PVC & PAC &
PVUAR & PAUAR & PVRL & PARL & VSR & ASR & HDF \\
\midrule
Spearman $\rho$ & 0.88 & 0.90 & 0.91 & 0.87 & 0.83 & 0.84 & 0.89 & 0.91 & 0.94 &
0.90 & 0.90 & 0.91 & 0.77 & 0.81 & 0.86 & 0.89 & 0.80 \\
\bottomrule
\end{tabular*}
\caption{Human alignment of selected evaluation metrics. We report Spearman
rank correlations between per-model human win rates and the corresponding
direction-adjusted benchmark scores.}
\label{tab:supp_human}
\end{table}

As shown in~\cref{tab:supp_human}, all 17 metrics show strong model-level human
alignment, with correlations ranging from 0.77 to 0.94. Prompt-based visual
continuity (\textit{PVC}) attains the highest correlation, whereas
visual and audio response latencies (\textit{PVRL} and \textit{PARL}) show
lower correlations than the corresponding achievement and fulfillment metrics.

\section{Checklist Construction Templates}

This section provides the fixed LLM prompt templates used to construct the
finalized visual and audio checklists and to decompose joint audio-video runtime
updates. Curly-brace placeholders denote case-specific content.

\subsection{Checklist Construction and Update Decomposition}

\begin{promptbox}{Visual Instruction Checklist Template}
# Task

Given a final audio-video generation prompt, construct a structured
checklist for Video Instruction Fulfillment (VIF).

The checklist contains four fields:

Scene
The environment or setting of the generated content, including specified
locations, surroundings, and important scene elements.

Visual Style
The specified visual style, artistic form, rendering style, or visual
medium.

Subjects
The central subjects specified in the prompt, including their number,
type or identity, and explicit visual attributes.

Activity
The primary activities specified in the prompt, including relevant
subject-subject or subject-object interactions.

# Instructions

- Generate one concise Yes/No question for each applicable field.
- Use only visual requirements explicitly stated in the prompt.
- Each question should be self-contained and objectively verifiable
  from the generated video.
- Closely related requirements within the same field may be combined.
- Do not introduce requirements that are not stated in the prompt.
- Phrase each question so that the expected answer is "Yes".
- If no applicable requirement exists for a field, return null.

Return strict JSON with no additional text.

# Output Format

{
  "video_instruction_following": {
    "scene": "<English question or null>",
    "visual_style": "<English question or null>",
    "subjects": "<English question or null>",
    "activity": "<English question or null>"
  }
}

Final generation prompt:

{generation_prompt}
\end{promptbox}

\begin{promptbox}{Audio Instruction Checklist Template}
# Task

Given a final audio-video generation prompt, construct a structured
checklist for Audio Instruction Fulfillment (AIF).

The checklist contains three fields:

Sound Source
The specified agent, entity, object, mechanism, or environmental source
responsible for producing a sound.

Sound Content
The specified audible events or sound content, including explicit
acoustic properties when applicable.

Temporal Relation
The specified temporal relationships among sounds, such as overlap,
simultaneity, separation, alternation, intermittency, or ordering.

# Instructions

- Generate one concise Yes/No question for each applicable field.
- Use only auditory requirements explicitly stated in the prompt.
- Each question should be self-contained and objectively verifiable
  from the generated audio-video content.
- Closely related requirements within the same field may be combined.
- Keep Sound Source focused on who or what produces the sound, and
  Sound Content focused on what is heard.
- Do not introduce requirements that are not stated in the prompt.
- Phrase each question so that the expected answer is "Yes".
- If no applicable requirement exists for a field, return null.

Return strict JSON with no additional text.

# Output Format

{
  "audio_instruction_following": {
    "sound_source": "<English question or null>",
    "sound_content": "<English question or null>",
    "temporal_relation": "<English question or null>"
  }
}

Final generation prompt:

{generation_prompt}
\end{promptbox}

\begin{promptbox}{Joint Audio-Video Update Splitter}
# Task

Split a Joint Audio-Video runtime update into a visual update and an audio
update.

# Instructions

- The visual update should contain the explicitly requested visual change.
- The audio update should contain the explicitly requested audio change.
- Preserve the original wording and semantic details whenever possible.
- Do not rewrite beyond what is necessary for the split or infer any new
  content.

Return strict JSON with no additional text.

# Output Format

{
  "visual_update": "<English visual update>",
  "audio_update": "<English audio update>"
}

Joint Audio-Video runtime update:

{runtime_update}
\end{promptbox}

\section{Example Benchmark Cases}

The following examples illustrate the prompt structure and temporal
dependencies used in the two evaluation tracks.

\subsection{Progressive Example}

A representative progressive case depicts a sailor crossing an open bay with a
border collie aboard. Its global prompt combines continuous steering and sail trimming with
wind-driven boat motion and a layered maritime soundscape.
\Cref{tab:supp_progressive_case} gives the complete prompt.

\begin{table}[H]
\centering
\small
\begin{tabular}{@{}rp{0.88\textwidth}@{}}
\toprule
Time & Prompt \\
\midrule
0--180 s &
A photorealistic compact sailing yacht crosses a breezy open bay beneath
broken afternoon clouds, with taut white sails, a weathered deck, rolling blue
water, and a distant low coastline; aboard, a windbreaker-clad adult sailor
continuously steers at the helm, monitors the wind, and trims the working lines
as the yacht heels and rises over the chop, while a distinct, alert
black-and-white border collie in a red flotation vest remains settled in the
cockpit and passively rides with the yacht's motion. The ratcheting winch and
creaking rigging remain prominent over the simultaneous sounds of rushing wind,
snapping canvas, and water striking the hull. \\
\bottomrule
\end{tabular}
\caption{Complete 180-second prompt example for the progressive track.}
\label{tab:supp_progressive_case}
\end{table}

\subsection{Interactive Example}

A representative interactive case follows a bread-baking process in a
photorealistic home kitchen. \Cref{tab:supp_interactive_case} gives the complete
prompt schedule. The heat
change at 60 seconds follows the dial adjustment at 30 seconds, closing the
window at 120 seconds responds to the rain introduced at 90 seconds, and the
crust color at 150 seconds reflects the higher oven setting established at
30 seconds.

\begin{table}[H]
\centering
\small
\begin{tabular}{@{}rp{0.88\textwidth}@{}}
\toprule
Time & Prompt \\
\midrule
0 s &
In a tidy photorealistic home kitchen, a stainless-steel wall oven steadily
bakes a pale loaf on its center rack beside an open, rain-clear window. A soft
refrigerator hum fills the pauses between occasional traffic whooshes outside. \\
30 s & A baker's hand rotates the oven temperature dial to a higher setting. \\
60 s & After the dial adjustment, the heat shimmer inside the oven intensifies. \\
90 s & Rain begins streaking the kitchen window. \\
120 s & The baker pulls the rain-streaked window shut with a firm click. \\
150 s & Under the higher oven setting, the loaf's pale crust turns golden brown. \\
\bottomrule
\end{tabular}
\caption{Complete 180-second prompt schedule example for the interactive track.}
\label{tab:supp_interactive_case}
\end{table}

\ifdefined\streamavcombined
\let\streamavenddocument\relax
\else
\clearpage
\bibliography{streamavbench2027}
\bibliographystyle{iclr2027_conference}
\def\streamavenddocument{\end{document}}
\fi
\streamavenddocument

\end{document}